\documentclass[preprint,showpacs,preprintnumbers,amsmath,amssymb]{revtex4}
\usepackage{graphicx,epsfig,dcolumn,bm,epic,eepic,float}%
\usepackage{amsmath}
\usepackage{latexsym}
\usepackage{color}
\usepackage{cancel}
\usepackage{makeidx,shortvrb,latexsym}
\usepackage{slashed}
\begin{document}
\unitlength 1 cm
\newcommand{\nn}{\nonumber}
\newcommand{\vk}{\vec k}
\newcommand{\vp}{\vec p}
\newcommand{\vq}{\vec q}
\newcommand{\vkp}{\vec {k'}}
\newcommand{\vpp}{\vec {p'}}
\newcommand{\vqp}{\vec {q'}}
\newcommand{\bk}{{\bf k}}
\newcommand{\bp}{{\bf p}}
\newcommand{\bq}{{\bf q}}
\newcommand{\br}{{\bf r}}
\newcommand{\bR}{{\bf R}}
\newcommand{\up}{\uparrow}
\newcommand{\down}{\downarrow}
\newcommand{\fns}{\footnotesize}
\newcommand{\ns}{\normalsize}
\newcommand{\cdag}{c^{\dagger}}
\title {Applying different
angular ordering constraints and $k_t$-factorization approaches  to the single inclusive hadron production in the ${e^ + }{e^ - }$ annihilation processes} 
\author{$M. \; Modarres$ }
\altaffiliation {Corresponding author, Email:
mmodares@ut.ac.ir, Tel:+98-21-61118645, Fax:+98-21-88004781.}
\author{$R. \; Taghavi$ }
\affiliation {Department of Physics, University of $Tehran$,
1439955961, $Tehran$, Iran.} \ \
\begin{abstract}
We study the differential cross section of the single inclusive ${e^ + }{e^ - }$ annihilation  to the hadrons via $\gamma$-production, in the different ${k_{t}}$-factorization frameworks. In
order to take into account the transverse momenta of the incoming partons, for the first time, we apply the  Kimber et al  (KMR) method to calculate the unintegrated parton fragmentation functions (UFFs) from the ordinary integrated one, i.e., the parton fragmentation functions (FFs), which satisfy the similar DGLAP evolution equations, such as  those of parton distribution functions (PDFs). Also, by utilizing the different angular ordering constraints the results corresponding to the Martin et al (MRW) in the leading order (LO) and the next-to-leading-order (NLO) are obtained. The LO sets of DSS library for  the input FFs is used.  The numerical results are compared with the experimental  data in the different energies which are reported by the different collaborations,   such as TASSO, AMY, MARK II, CELLO, DELPHI, SLD, ALEPH and Belle with  the other   QCD+fragmentation models such as Pythia 6.4 and  8.2 parton showers. The behaviors of the normalized  differential cross sections and the multiplicity versus the "transverse momentum" ($p_\bot$)  are discussed. The final  results demonstrate that the KMR and MRW UFFs give a good description of data and there is not much significant difference between the above three schemes.  On the other hand, our results become closer to the data for the lower values of $p_\bot$     and  the higher values of center of mass energies.
 \end{abstract} 

\pacs{12.38.Bx, 13.85.Qk, 13.60.-r
\\ \textbf{Keywords:}  unintegrated fragmentation function,
${k_{t}}$-factorization, electron-positron pair annihilation, transverse momentum.
\\
\\ (In the on-line journal, the figures are colored)} \maketitle

\section{Introduction}
The discovery of the partonic structure of hadrons based on quarks and gluons is one of the most interesting topics in the theoretical and experimental high energy physics. The parton distribution functions (PDFs) represent the densities of these fundamental particles which initially depend on the Bjorken variable $x$ and the hard scale ${\mu ^2}$ by DGLAP evolution equations \cite{DGLAP1, DGLAP2, DGLAP3, DGLAP4}. However, the experimental data of hadron-hadron colliders show that the significant information is embedded in the transverse momentum of the initial hadron constituents. So the important inputs are the unintegrated     parton
distribution functions (UPDFs). The UPDFs can be interpreted as a number densities of partons that are carrying a fraction $x$ of the momentum of parent hadron with the transverse momentum $k_t$ at the  hard scale ${\mu ^2}$. These transverse dependent functions   extensively were investigated in the Drell-Yan and  the semi-inclusive deep inelastic scattering (SIDIS) processes and unlike the collinear ones, they are still highly debated subjects \cite{lie, mypaper2, aybat}. Theoretically, various methods are utilized to generate these
fundamental quantities, and among them, the Kimber et al \cite{KMR} and Martin et al \cite{MRW} formalisms are more simplistic ways to describe these UPDFs. The general behavior of these
prescriptions  was investigated in the references
\cite{10,11,12,13,14}.

Of equal importance is the hadronization mechanism of the generation of mesons and baryons from partons. To reach a comprehensive description of these processes the fragmentation functions (FFs) are required \cite{OWENS1,Owens1,Owens2,FF1,Metz}. These non-perturbative fundamental quantities mean the probability of carrying the light-cone fraction $z$ of the fragmenting parent parton by hadron $H$, and can be measured in SIDIS and single- or double-inclusive hadron production in the electron-positron annihilation processes. While FFs are necessarily coupled to the PDFs in SIDIS, the single inclusive annihilation (SIA) provides a golden channel and a cleanest electromagnetic probe to study FFs, because there is no contribution from hadronic effects in the initial states \cite{clean1}. At the first order and center of mass (CM) energies below the $Z^0$ mass, this process can be interpreted as $
{e^ + }{e^ - }   \to HX$, via a single virtual photon, which can subsequently fragment into the hadrons ($H$). The direction of
the fragmenting back to back $q{\bar q}$ pair is identified by the jet
axis resulting from each ${e^ + }{e^ - }$ scattering, and the detected $p_\bot$ represents a direct measurement of the transverse momentum of the final hadron with respect to the fragmenting parent parton.

Some data upon polarized Collins FFs were achieved by BABAR collaboration \cite{BABAR1,BABAR2} and some literature on this subject are presented \cite{collins1,collins2,collins4}. But, a little experimental information exists on the  unpolarized transverse momentum dependent FFs. Although a thorough knowledge of these functions
would be of fundamental importance for studying the transverse motion of hadrons, because of lack of data on ${p_\bot }$ distribution of  $
{e^ + }{e^ - }$ unpolarized cross sections, limited studies performed over these functions \cite{f1,f2,f3,f4,f5,mypaper1,BR1,BR2,f7}. However, in this paper, we concentrate on the data for single-inclusive  hadron cross section in the  ${e^ + }{e^ - }$ annihilation process, ${e^ + }{e^ - }   \to HX$, from TASSO collaboration at PETRA (DESY) \cite{TASSO1,TASSO2}. The advantage of these data set, which are integrated over $z$ with a small average value of $z_H$, $\langle z_H \rangle$, is delivering measurements
at the different CM energies. They provide the differential cross sections in terms of
$p_\bot$, normalized to the fully inclusive cross section which has interesting features to be studied under the $k_t$-factorization scheme. In this work, the cross section data as a function of $p_\bot$ distributions, integrated over the energy fraction $z$ of the detected hadron $H$, for all charged particles production in the different CM energies between 14 and 44 GeV are considered. Moreover, we also consider the MARKII \cite{mark2}, AMY  \cite{AMY} and CELLO \cite{CELLO} collaboration data collected at the SLAC storage ring PEP, the KEK collider TRISTAN and at the PETRA, respectively. Also, the ${e^ + }{e^ - }$ unpolarized cross sections are discussed for Pythia 6.4 and Pythia 8.2 parton showers \cite{R2 new 2} in comparison with those  of DELPHI \cite{R2 new 3}, SLD
\cite{R2 new 4} and ALEPH \cite{R2 new 5} collaborations data at CM energy 91 GeV. 
%\begin{table}
%\centering
%{\scriptsize
%\caption{Low-energy single inclusive ${e^ + }{e^ - }$ annihilation experimental data: This table includes respectively from top to bottom the reference addresses of data, number of data points, the values of the center of mass and beam  energy, the type of beams and targets, the range of  invariant mass bins, ${p_\bot}$ range, the used observables, and finally the amount of systematic uncertainty for each experiment.}
%\begin{tabular}{ccccccccccc}
%\hline
%\hline
% & Experiment  &&& c.m.energy (GeV)  &&&Reference&&&$\langle z_H \rangle$ \\ 
%\hline
%&TASSO&&& $ 14 $&&&\cite{tasso2}&&& $ 0.13$   \\
% &&&& $ 22 $&&&\cite{tasso2} &&& $ 0.11 $    \\
%& &&& $ 35 $ &&&\cite{tasso2}&&& $0.09 $    \\
%& &&& $ 44 $ &&&\cite{tasso2}&&& $0.08 $     \\
%&MARKII &&& $ 29 $&&&\cite{mark2} &&& $0.09 $ \\
%&AMY &&& $ 52-57 $&&&\cite{AMY} &&& $- $   \\
%&CELLO &&& $ 34 $&&&\cite{CELLO}&& & $- $   \\

%\hline
%\hline
%\end{tabular}}

%\end{table}

Although these data are old and limited to the ${p_\bot }$ distribution, they represent extremely valuable and unique information of a direct measurement of intrinsic transverse momenta of the final hadrons with respect to the fragmented parent parton. These data in the non-perturbative region, ${p_ \bot }<1$ GeV, are phenomenologically studied by   Boglione et al \cite{mypaper1,BR1,BR2}, considering two functional forms i.e. the gaussian and the power-law, as models for fitting to reproduce the behavior of data at small $p_\bot$. There is also
 the newer data from Belle collaboration \cite{BR3}. These data provide the unpolarized cross sections of charged pions and kaons based on z, ${p_ \bot }$ and event shape variable ($d^3\sigma /dzdp_\bot  dT$) in the $\sqrt s$=10.58 GeV  Belle Collaboration \cite{BR3}. In this data sets the transverse momentum of the produced hadron is calculated relative to the thrust axis ${\bf \hat{n}}$ which maximizes the event-shape variable thrust T: 
 $$
 T \stackrel{\text{max}}={\sum\limits_h |{\bf P}_{h}^{CMS}\cdot{\bf \hat{n}}|\over \sum\limits_h |{\bf P}_{h}^{CMS}|},
 $$
in which the sum runs over all detected particles, and the momentum of hadron $h$ in the center-of-mass system (CMS), denotes by $P^{CMS}_h$. It is shown in the reference \cite{BR3} that $ uds$ and charm events has a peak at high thrust values. That is why in this work, our results will be displayed in the $0.85 <T<0.9$ thrust bin. We also select high z bins data sets, because our perturbative formulisems are valid for $p_\bot>1$ GeV. 
 
In this work, we intend  to
constrain our analysis to the region of ${p_ \bot }>1$ GeV in which there are perturbative effects. We apply, for the first time, the KMR and MRW formalisms  in the leading and next-to-leading order to test the capability of these procedures  in obtaining the UFFs, $D(z,p_\bot,\mu^2)$.  It is essential to emphasize  that our goal here is not so much the determination of UFFs, which would require all possible processes, but rather to explore, the application of the KMR and MRW methodology for finding UFFs. So, we restrict ourselves to the lowest order of QCD, and neglect all terms related to $\alpha _s({\mu^2})$ in the differential cross section calculations. 

The results are compared with distributions generated by QCD+fragmentation model programs via Monte Carlo techniques such as the leading-logarithmic parton shower (Lund PS) \cite{LundPS},  the second-order matrix-element calculation (Lund ME) \cite{LundME}, and the model of the Gottschalk and Morris (CALTECH II) \cite{caltech2} at the parton level.  The main approach of all these programs is utilizing a model with a few free parameters in the fitting to the data processes. The important features of these models are briefly discussed in the references \cite{AMY, mark2}. Beside these parton showers, there is also possibility of  comparison of the result with those of Pythia 6.4 and Pythia 8.2 \cite{R2 new 2}.   

 The organization of our paper is as follows: In section II, we review briefly the basic formulas of the cross section of ${e^ + }{e^ - }$ annihilation into hadrons, 
${k_{t}}$-factorization approach, and the KMR methodology to construct the UPDFs and UFFs. In section III we present the numerical results and discussions. Finally, we summarize our conclusions in section IV.

\section{The FORMALISM}
\subsection{The cross section and the fragmentation functions}
In this section, we present some theoretical aspects of the cross section of  the single inclusive hadron production in ${e^ + }{e^ - }$ annihilation process into a single hadron $H$,\
\begin{equation}
{e^ + }{e^ - } \to \gamma  \to HX.
\end{equation}
by considering unpolarized fragmentation function. The cross section for such a process by including the transverse momentum can be  typically written  in the following form:
 \begin{equation}
\frac{1}{{{\sigma _{tot}}}}\frac{{d{\sigma ^H}}}{{dz{d^2}{{\vec p}_ \bot }}} = \frac{1}{{\sum\limits_q {{e_q}^2} }}  {[2F_1^H(z,{p_ \bot };{\mu^2}) + } F_L^H(z,{p_ \bot };{\mu^2})],\label{sigma}
\end{equation}
where ${e_q}$ is the charge of each quark flavor and the sum runs over all active quark-antiquark flavors. The energy ${E_H}$ of fragmented hadron with respect to the beam energy ${\sqrt s\over 2}$ is presented by the parameter  $z = 2{p_H}.q/{\mu^2} = 2{E_H}/\sqrt s $ which, in the ${e^ + }{e^ - }$   CM frame, is interpreted as the momentum fraction of the parent quark
carried by the produced hadron. Details on the unpolarized “time-like” structure functions $F_1^H$ and $F_L^H$ in the equation (2) can be
found in the reference \cite{Altarelli,f1fh}. The total cross section for the ${e^ + }{e^ - }$ annihilation to hadrons is presented by:
\begin{equation}
{\sigma _{tot}}(\mu^2) = \sum\limits_q {{e_q}^2} {\sigma _0}[1 + \frac{{{\alpha _s}({\mu^2})}}{\pi }]+\mathcal{O}(\alpha _s^2),
\end{equation}
where ${\sigma _0} = \frac{{4\pi {\alpha ^2}}}{{3s}}$ in which $\alpha=e^2/4\pi$ denotes the electromagnetic fine structure constant, and in the leading order in $\alpha_s$, we have ${\sigma _{tot}} = \frac{{4\pi {\alpha ^2}}}{{3s}}\sum\limits_q {{e_q}^2}$ .

The structure functions ${F_1^H}$ in the leading order accuracy is given by:
\begin{equation}
2F_1^H(z,{p_\bot};{\mu^2}) = \sum\limits_q {{e_q}^2}   [D_q^H(z,{p_\bot};{\mu^2}) + D_{\bar q}^H(z,{p_\bot};{\mu^2})] \hfill \\
 .  \hfill \\
\end{equation}
%\begin{equation}
%2F_1^H(z,{p_\bot};{Q^2}) = \sum\limits_q {{e_q}^2} \{ [D_q^H(z,{p_\bot};{Q^2}) + D_{\bar q}^H(z,{p_\bot};{Q^2})]\nonumber\\
%\end{equation}
%\begin{equation}
 %+ \frac{{{\alpha ^2}({Q^2})}}{{2\pi }}[C_q^1 \otimes (D_q^H + D_{\bar q}^H) + C_g^1 \otimes D_g^H](z,{p_\bot};{Q^2})\} .
%\end{equation}
and $F_L^H$ has not any term in the leading order.   
In this equation, the ${{D_q^H}(z,{p_\bot};\mu^2)}$ is the ordinary unpolarized single-hadron FFs.
In our analysis, we restrict ourselves to the leading order approximation in which the ${p_\bot}^2/{\mu^2} \ll 1$ limitation is applied. With these considerations, we may simply  write the differential cross section formula as following :
\begin{equation}
\frac{1}{{{\sigma _{tot}}}}\frac{{d{\sigma ^H}}}{{dz{d^2}{{\vec p}_\bot}}} =\frac{1}{{\sum\limits_q {{e_q}^2} }} \sum\limits_q {{e_q}^2} [D_q^H(z,{p_\bot};{\mu^2}) + D_{\bar q}^H(z,{p_\bot};{\mu^2})].
\end{equation}
 
After integrating the  equation (5) over $z$ and also considering ${d^2}{{\vec p}_\bot} = 2\pi {p_\bot}d{p_\bot}$, we have the final formula   for the differential cross sections in the LO with respect to $p_\bot$ :
\begin{equation}
\frac{1}{{{\sigma _{tot}}}}\frac{{d\sigma^H }}{{d{p_\bot}}} = 2\pi {p_\bot}\frac{1}{{\sum\limits_q {{e_q}^2} }}\int {\sum\limits_q {{e_q}^2[D_q^H(z,{p_\bot};{\mu^2}) + } D_{\bar q}^H(z,{p_\bot};{\mu^2})]dz},
\end{equation}
and 
\begin{equation}
\frac{1}{{{\sigma _{tot}}}}\frac{{d\sigma^H }}{{dz}} = 2\pi \frac{1}{{\sum\limits_q {{e_q}^2} }}\int {\sum\limits_q {{e_q}^2[D_q^H(z,{p_\bot};{\mu^2}) + } D_{\bar q}^H(z,{p_\bot};{\mu^2})]{p_\bot}d{p_\bot}}.
\end{equation}
\subsection{The KMR and MRW prescriptions, UPDFs and UFFs}
In the KMR \cite{KMR} method by starting from the DGLAP evolution equation, and performing $k_t$ factorization prescriptions, we obtain  the UPDF of each parton which depends on the transverse momentum $k_t$, the
fractional momentum $x$ at hard scale $\mu^2$ as:
 \begin{equation}
f_a(x,k_t,\mu^2) = T_a(k_t,\mu^2)\sum_{b} \left[
{\alpha_S(k_t^2) \over 2\pi} \int^{1-\Delta}_{x} dz P_{ab}^{(0)}(z)
b\left( {x \over z}, k_t^2 \right) \right] , \label{eq1}
    \end{equation}
where  the familiar double logarithmic Sudakov survival factor ${T_a(k_t,\mu^2)}$ is
  \begin{equation}
T_a(k_t,\mu^2) = exp \left( - \int_{k_t^2}^{\mu^2} {\alpha_S(\kappa_t^2)
\over 2\pi} {d\kappa_t^{2} \over \kappa_t^2} \sum_{b} \int^{1-\Delta^\prime}_{0} dz'
P_{ba}^{(0)}(z') \right). \label{eq2}
    \end{equation}
where   ${P_{ab}^{(0)}}(z)$ ($b=q, \bar{q}$  and $ g)$
denotes the usual LO splitting functions and    ${b\left( {x \over z}, k_t^2 \right)}$  are the LO
PDFs (the Sudakov form factor becomes equal one for $k_t>\mu$). In this formula, the
angular-ordering constraint (AOC) \cite{CCFM1,CCFM2,CCFM3,CCFM4,CCFM5,44,45}, $\Delta$ ($\Delta^\prime$), is applied in the upper limit of the integration, which is an infrared cutoff to prevent the soft gluon singularities arise from the splitting
functions and defined as $\Delta= {k_t \over \mu + k_t}$ ($\Delta^\prime= {\kappa_t \over \mu + \kappa_t}$), which constrains the $k_t$ ordering.
 
\begin{figure*}[htp] 
\begin{center}
\begin{tabular}{cc}
{\includegraphics[width=100mm,height=20mm]{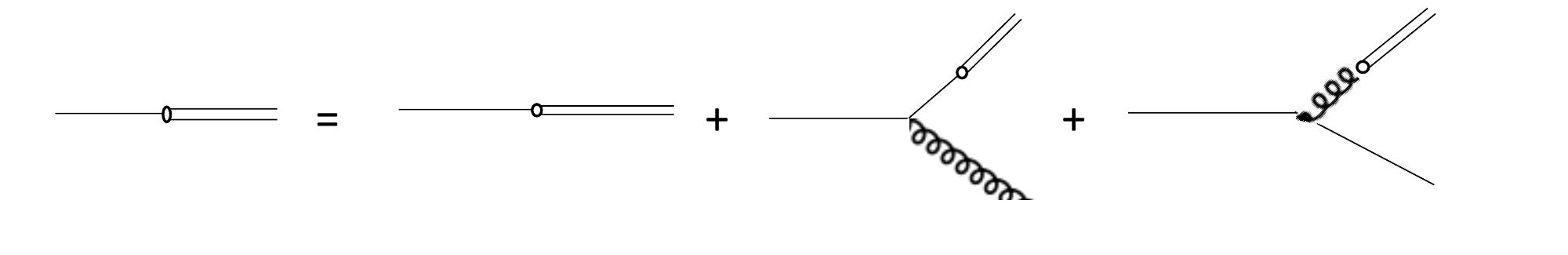}}  
 \end{tabular}
\caption{The graphical representation  of the equation (\ref{eq3}). } \label{fig:0}
\end{center}
\end{figure*}

Similarly, we can obtain a formula for the quark UFFs by starting from the complete (leading order) DGLAP evolution equation for the quark FFs in terms of qurak and gluons, see the figure \ref{fig:0} (a similar equation can be written for the anti-quarks, through out of this report):
\begin{equation}
{\partial D_{q}^{H}(z,\mu^2) \over{\partial ln \mu^2} } =
{\alpha_S(\mu^2) \over 2\pi} \int^{1}_{x}{dx \over x}\left[{{P_{qq}(x) D_q^H\left( {z \over x}, \mu^2 \right) } +{P_{gq}(x) D_g^H\left( {z \over x}, \mu^2 \right) }}\right]
 , \label{eq3}
    \end{equation}
Note that in this formula, according to the reference \cite{OWENS1,Metz}, the functions $P_{gq}$ are interchanged in comparison with that of parton evolution equations. The relevant splitting kernels are:
\begin{equation}
P_{qq}(x)=C_F\left[{{1+x^2}\over{(1-x)_+}} + {{3}\over {2}} \delta (1-x) \right].
    \end{equation}
\begin{equation}
P_{gq}(x)=C_F{{1+(1-x^2)}\over x} .
    \end{equation}
By inserting these splitting kernels and using the $plus$ prescription in a straightforward way, one could have the LO DGLAP equation evaluated at a scale $k_t$:
\begin{equation}
{\partial {\cal D}_{q}^{H}(z,k_t^2) \over{\partial ln k_t^2} } =
{\alpha_S(k_t^2) \over 2\pi}\{\sum_{a} \int^{1-\Delta}_{z}{P_{aq}(x) {\cal D}_a^H\left( {z \over x}, k_t^2 \right) {dx}} -{\cal D}_q^H\left( {z }, k_t^2 \right)\sum_{a}\int^{1-\Delta}_{x}{ P_{qa}(z') dz'}\}
 . \label{eq4}
    \end{equation}
Here, $P_{aq}$  refer to the unregulated LO DGLAP splitting kernels and  $ {\cal D}_{i}^{H}(z,k_t^2)=z{ D}_{i}^{H}(z,k_t^2)$ ($i=q, \bar{q}$  and $ g)$. The two terms on the right
hand side correspond to real  and virtual emission respectively. The virtual contributions may be resummed to all orders by the Sudakov form factor,
\begin{equation}
T_q(k_t,\mu^2) = exp \left( - \int_{k_t^2}^{\mu^2} {\alpha_S(\kappa_t^2)
\over 2\pi} {d\kappa_t^{2} \over \kappa_t^2} \sum_{b} \int^{1-\Delta}_{0} dz'
P_{qb}^{(0)}(z') \right). \label{eq5}
    \end{equation}
which is the   survival probability that hadron $H$ with transverse momentum $k_t$ remains untouched in the evolution, up to the factorization scale $\mu$. Therefore, the  UFFs become dependent on
the two scales, $k_t^2$ and $\mu^2$ in the $last$ $step$ of the evolution. So, in the $k_t$-factorization frame-work, the UFFs have the following forms:
 \begin{equation}
D_q^H(z,k_t,\mu^2) = T_q(k_t,\mu^2)\sum_{b=q,g} \left[
{\alpha_S(k_t^2) \over {2\pi k_t^2}} \int^{1-\Delta}_{z} dz' P_{bq}^{(0)}(z')
{\cal D}_b^H\left( {z \over z'},k_t^2 \right) \right] , \label{eq6}
    \end{equation}
where  ${{\cal D}_b^H\left( {z \over z'}, k_t^2 \right)}$  are the collinear, unpolarized quarks and gluons  FFs. We use the LO set of DSS \cite{DSS}.  
Note that  in general we set $k_t=p_\bot/z$ \cite{mypaper1} for the UFFs to calculate   different differential cross sections.

By applying the AOC only on the terms which include the on shell gluon emissions for the quarks  and gluons, we have the LO-MRW UFFs for quarks in the following form:
$$
{ D}_q^{H,LO}(z,k_t,\mu^2)= T_q(k_t,\mu^2) {\alpha_S(k_t^2) \over
{2\pi k_t^2}} \int_z^1 dz' \left[ P_{qq}^{(0)}(z') {z \over z'} D^H_q \left( {z
\over z'} , k_t^2 \right) \Theta \left( {\mu \over \mu +k_t}-z'
\right) \right.
    $$
\begin{equation}
\left. + P_{gq}^{(0)}(z') {z \over z'} D_g \left( {z \over z'} , k_t^2
\right) \right],
    \end{equation}
with
  \begin{equation}
T_q(k_t,\mu^2) = exp \left( - \int_{k_t^2}^{\mu^2} {\alpha_S(\kappa_t^2)
\over 2\pi} {d\kappa_t^{2} \over \kappa_t^2} \sum_{b} \int^{1-\Delta}_{0} dz'
P_{qb}^{(0)}(z') \right). \label{eq5}
    \end{equation}
  By expanding the MRW formalism to the NLO level, we have:
    $$
{  D}_q^{H,NLO}(z,k_t,\mu^2)= \int_z^1 dz' T_q \left(k, \mu^2 \right) {\alpha_S(k^2) \over {2\pi k_t^2}}
    \sum_{b=q,\bar{q},g} \tilde{P}_{bq}^{(0+1)}(z')
    $$
    \begin{equation}
\times D_b^{H,NLO} \left( {z \over z'} , k^2 \right) \Theta \left(
1-z'-{k_t^2 \over \mu^2} \right),
    \label{eq11}
    \end{equation}
where $k^2={k_t^2 \over
(1-z')}$.
In the above formula, the Sudakov form factor is defined as:
  \begin{equation}
T_q(k,\mu^2) = exp \left( - \int_{k^2}^{\mu^2} {\alpha_S(\kappa_t^2)
\over 2\pi} {d\kappa_t^{2} \over \kappa_t^2} \int^1_0 dz' z' \left[
\tilde{P}_{qq}^{(0+1)}(z') + \tilde{P}_{qg}^{(0+1)}(z') \right]
\right),
        \end{equation} 
The higher order  splitting functions
are presented in the appendix A.
\section{Numerical results and discussions}
In this section, we intend to present the kinematic and theoretical aspects of our calculations. First, we calculate the  KMR UFF based on the ${k_{t}}$-factorization scheme by applying exactly analogous steps hold for the UPDF \cite{KMR} which was developed in the section II. The similar perturbative calculation for both LO-MRW and NLO-MRW FF are also implemented. It is important, however, to point out that the crucial constraint for instructing any new UFF is the normalization relation,
\begin{equation}
D^H_{q}(z,\mu^2) \simeq \int^{\mu^2} {{dp_\bot^2} }
D^H_{q}(z,p_\bot,\mu^2).\label{NOR}
\end{equation}

In this article, we attempt to extract information about the perturbative evolution region. So, we
restrict  our analysis to the region of $p_\bot> 1.0$ GeV. Moreover, we vary the scale $\mu$ between $\mu/2$ and $2\mu$ to assess the uncertainty in the perturbative calculation.

\begin{figure*}[htp]
  \begin{center}
    \begin{tabular}{cc}
      {\includegraphics[width=80mm,height=60mm]{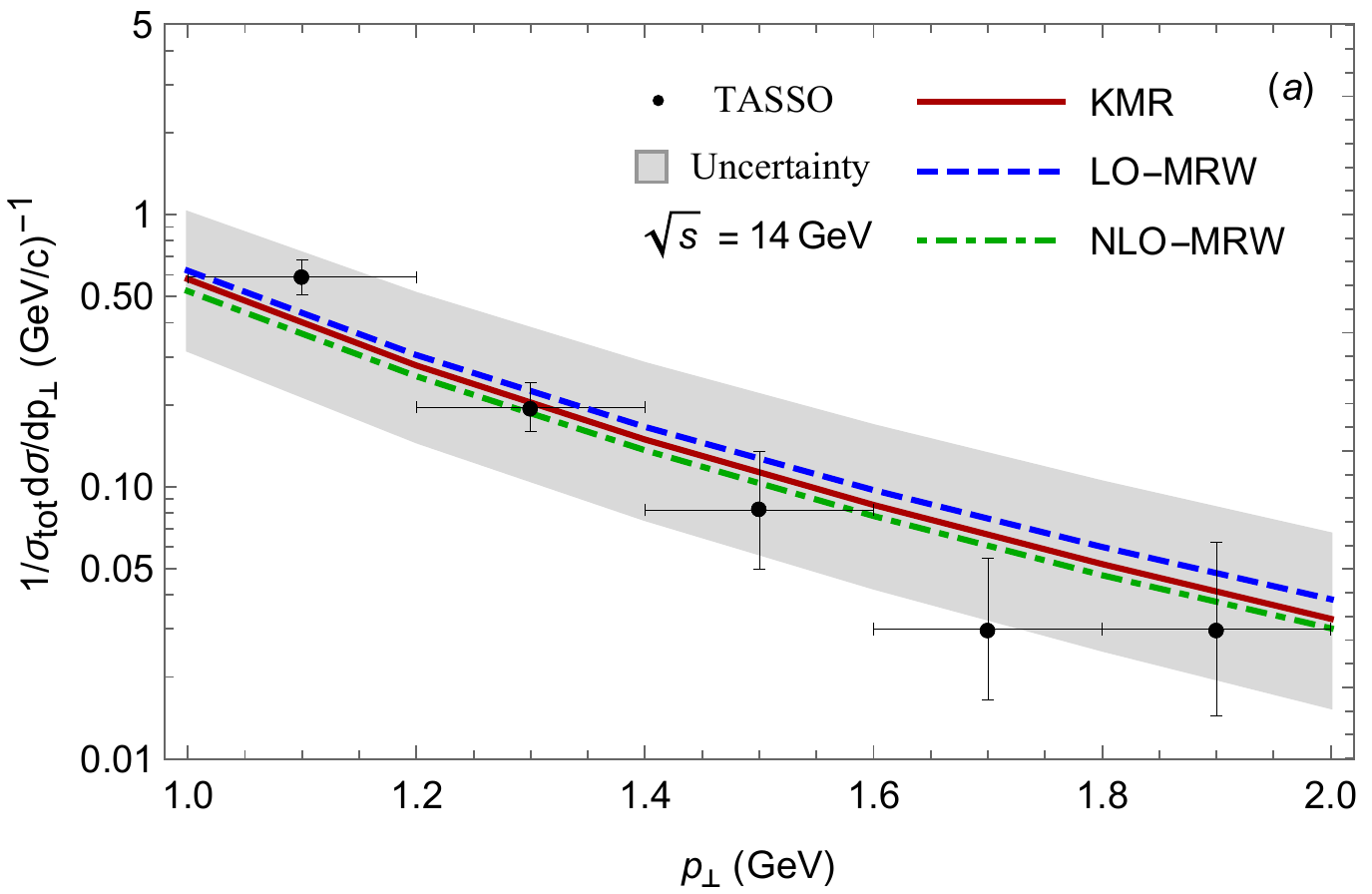}} &
       {\includegraphics[width=80mm,height=60mm]{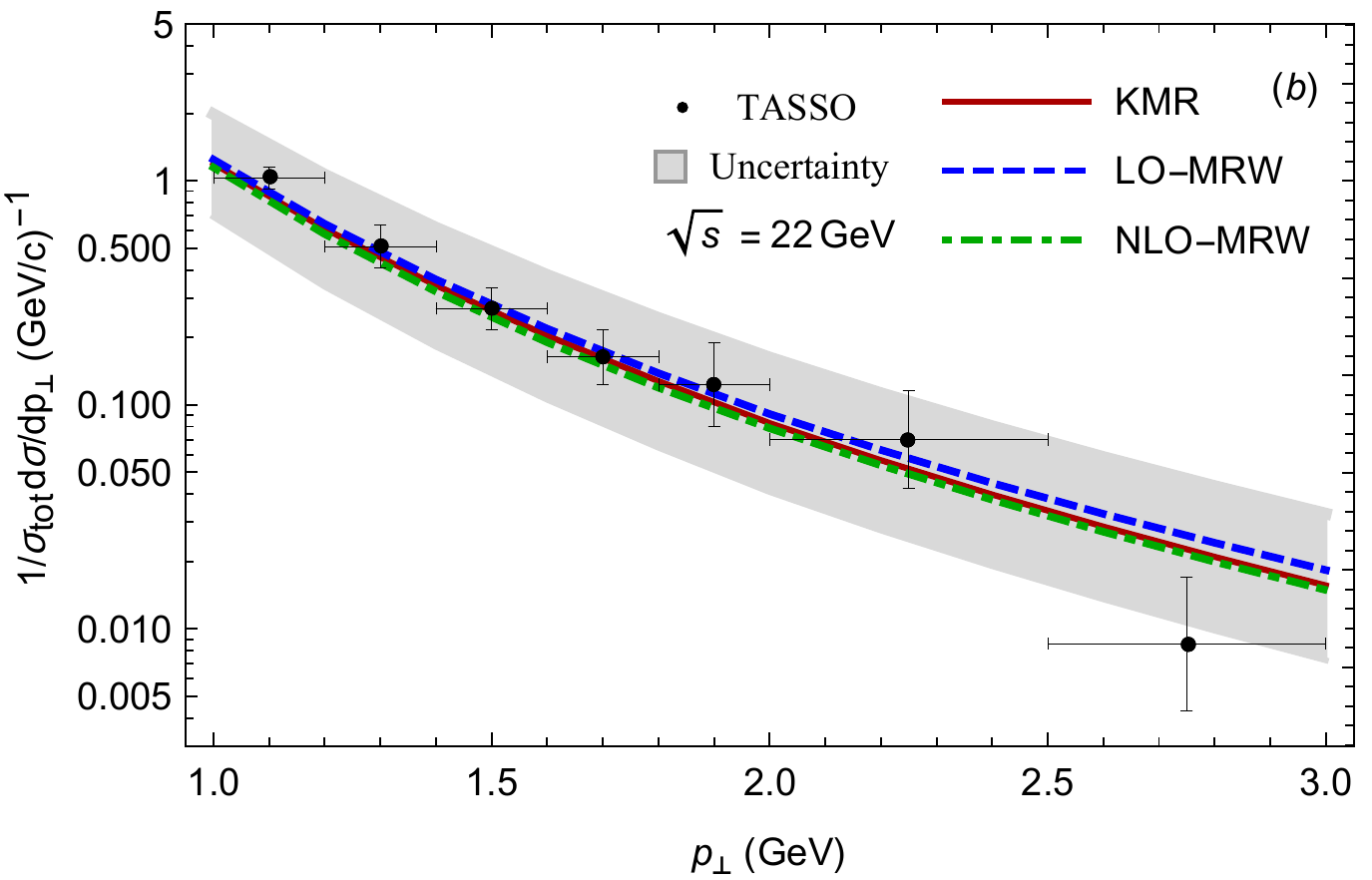}} \\
       {\includegraphics[width=80mm,height=60mm]{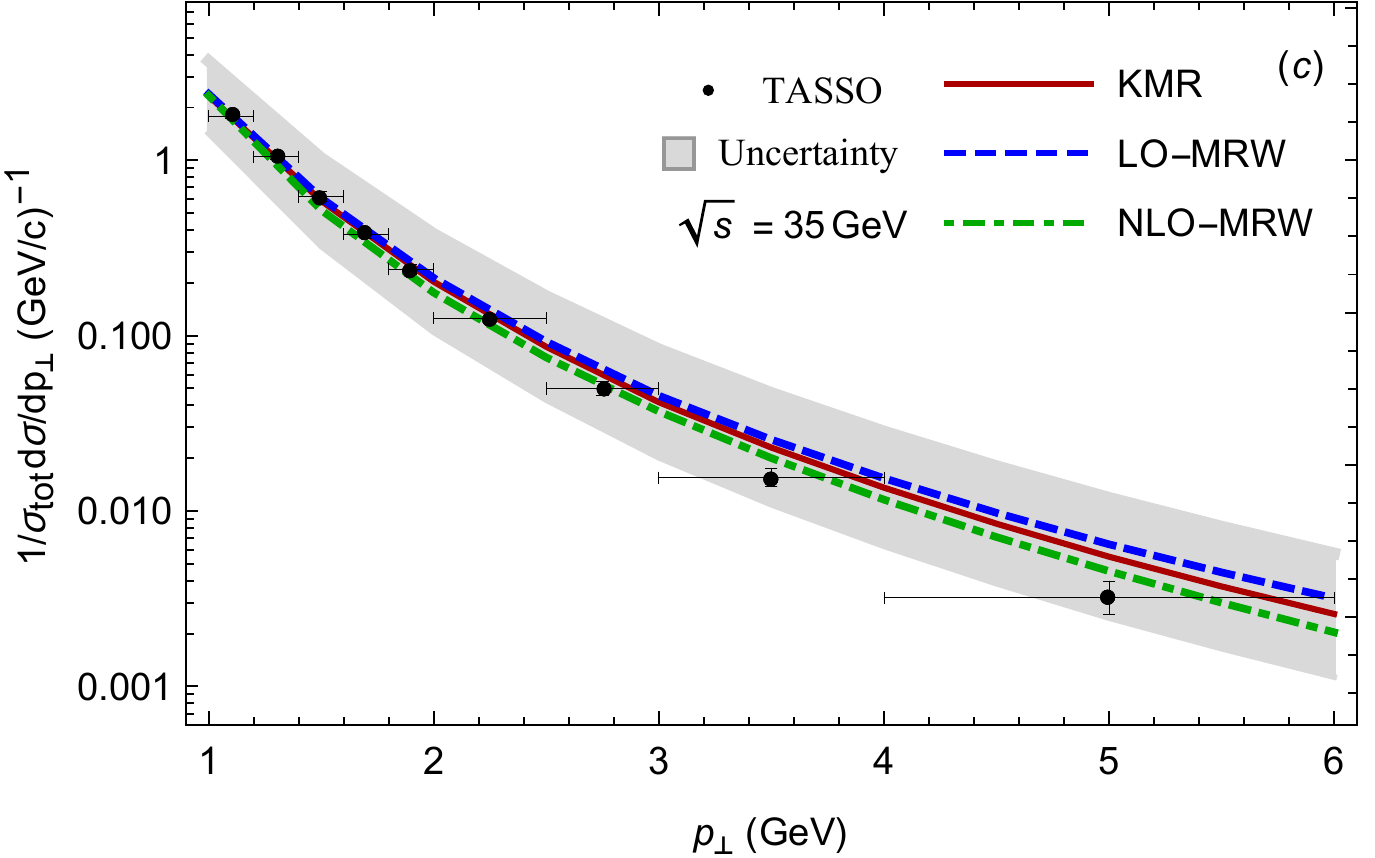}} &
       {\includegraphics[width=80mm,height=60mm]{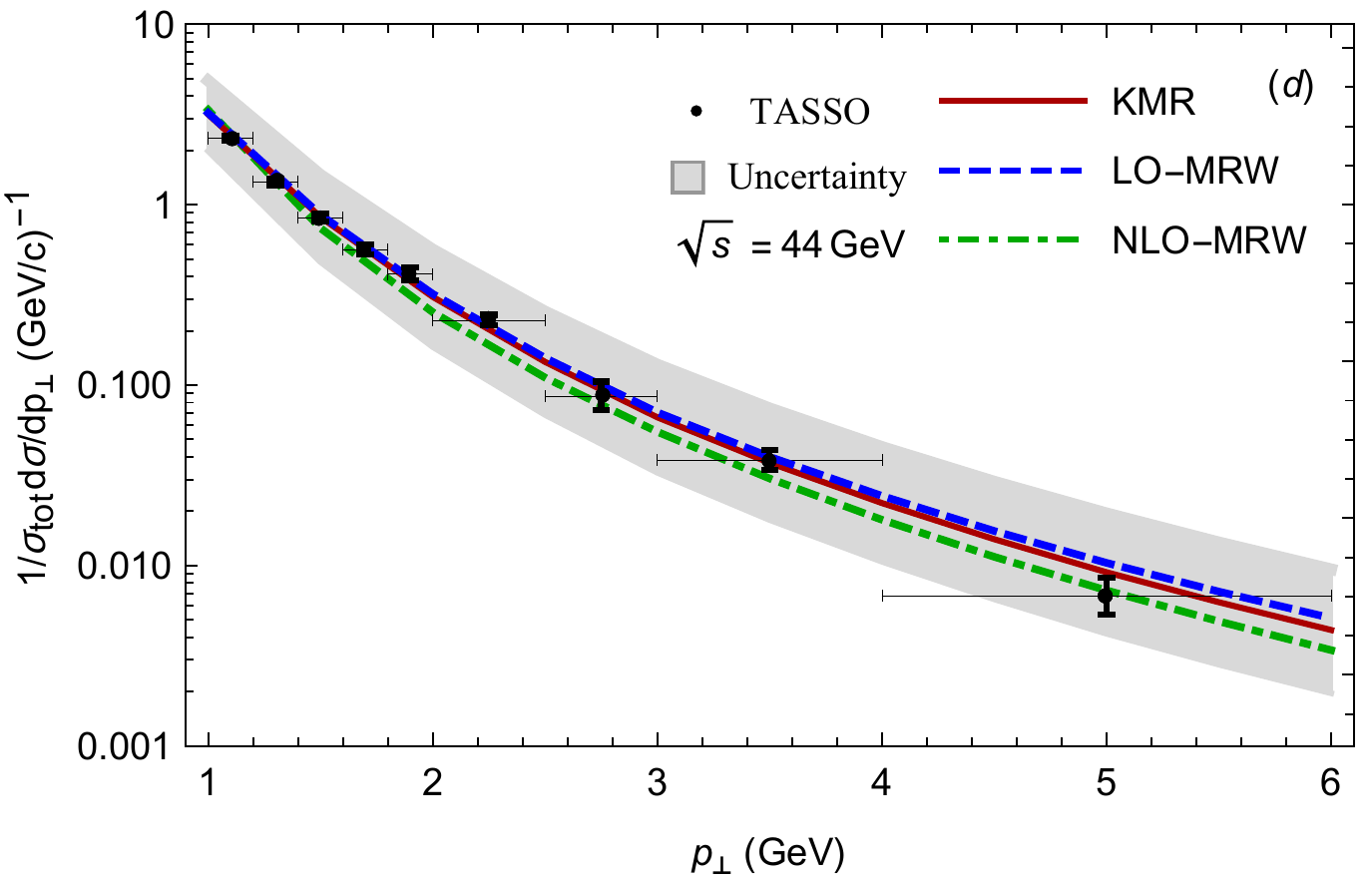}} \\
     \end{tabular}
\caption{ The normalized differential cross sections
($1/\sigma_{tot}$)$d\sigma/dp_\bot$ with respect to  $p_\bot$  compared  to the experimental data  of TASSO \cite{TASSO2} at the different CM energies. The shaded uncertainty grey bands are belong to the KMR prescription.\label{fig:1}}
      \end{center}
\end{figure*}

\begin{figure*}[htp]
  \begin{center}
    \begin{tabular}{cc}
      {\includegraphics[width=80mm,height=60mm]{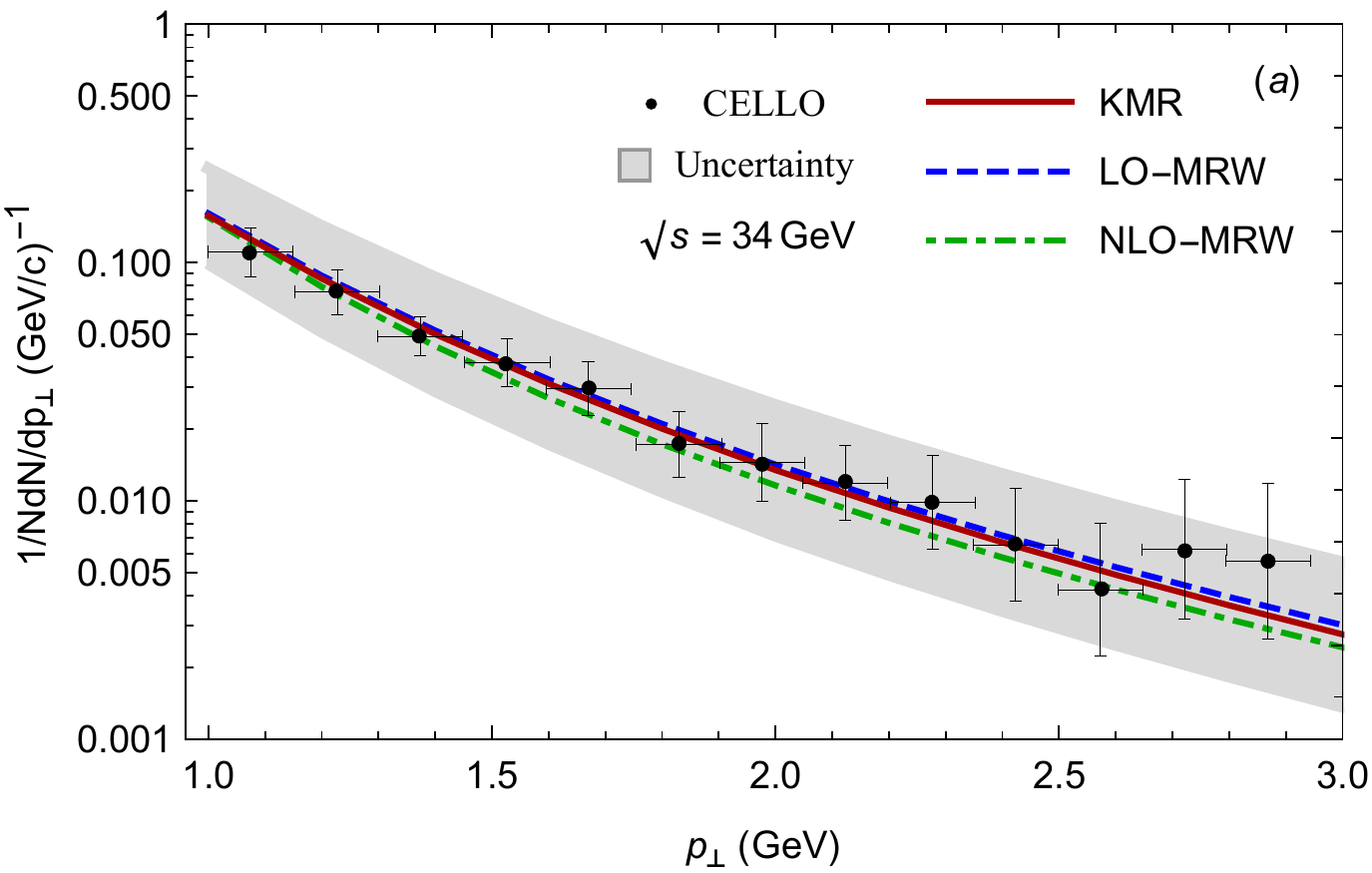}} &
       {\includegraphics[width=80mm,height=60mm]{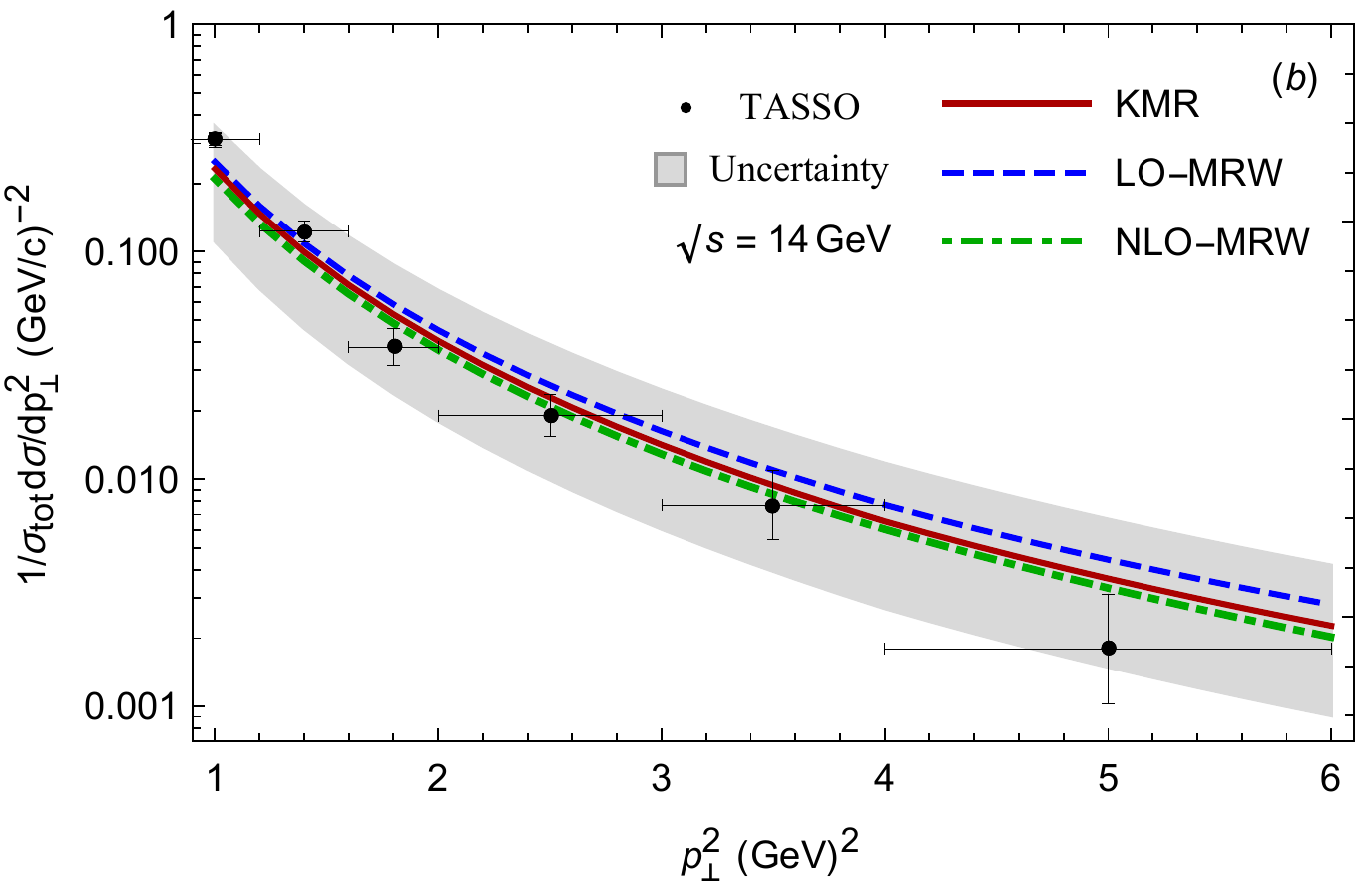}} \\
       {\includegraphics[width=80mm,height=60mm]{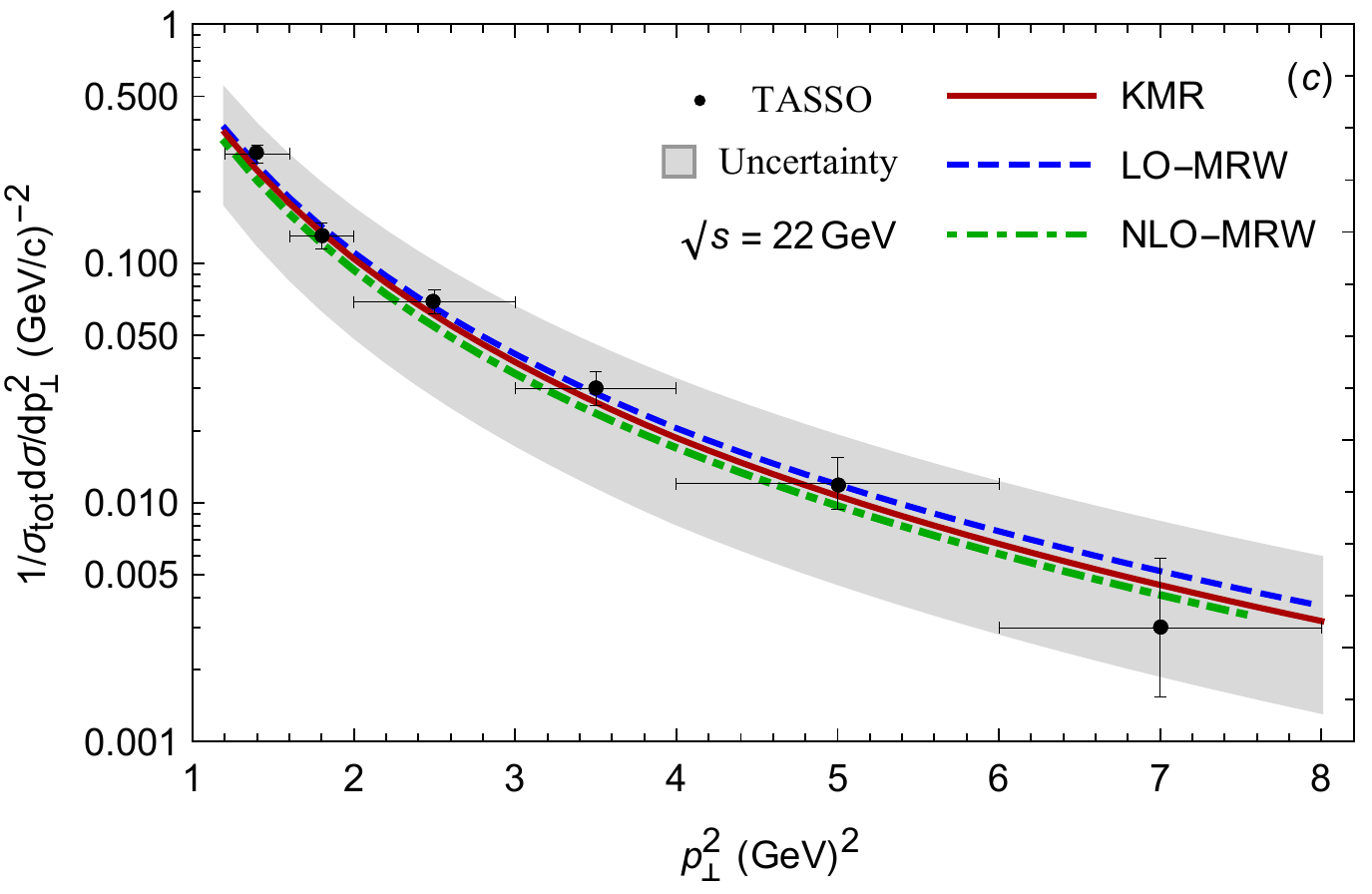}} &
       {\includegraphics[width=80mm,height=60mm]{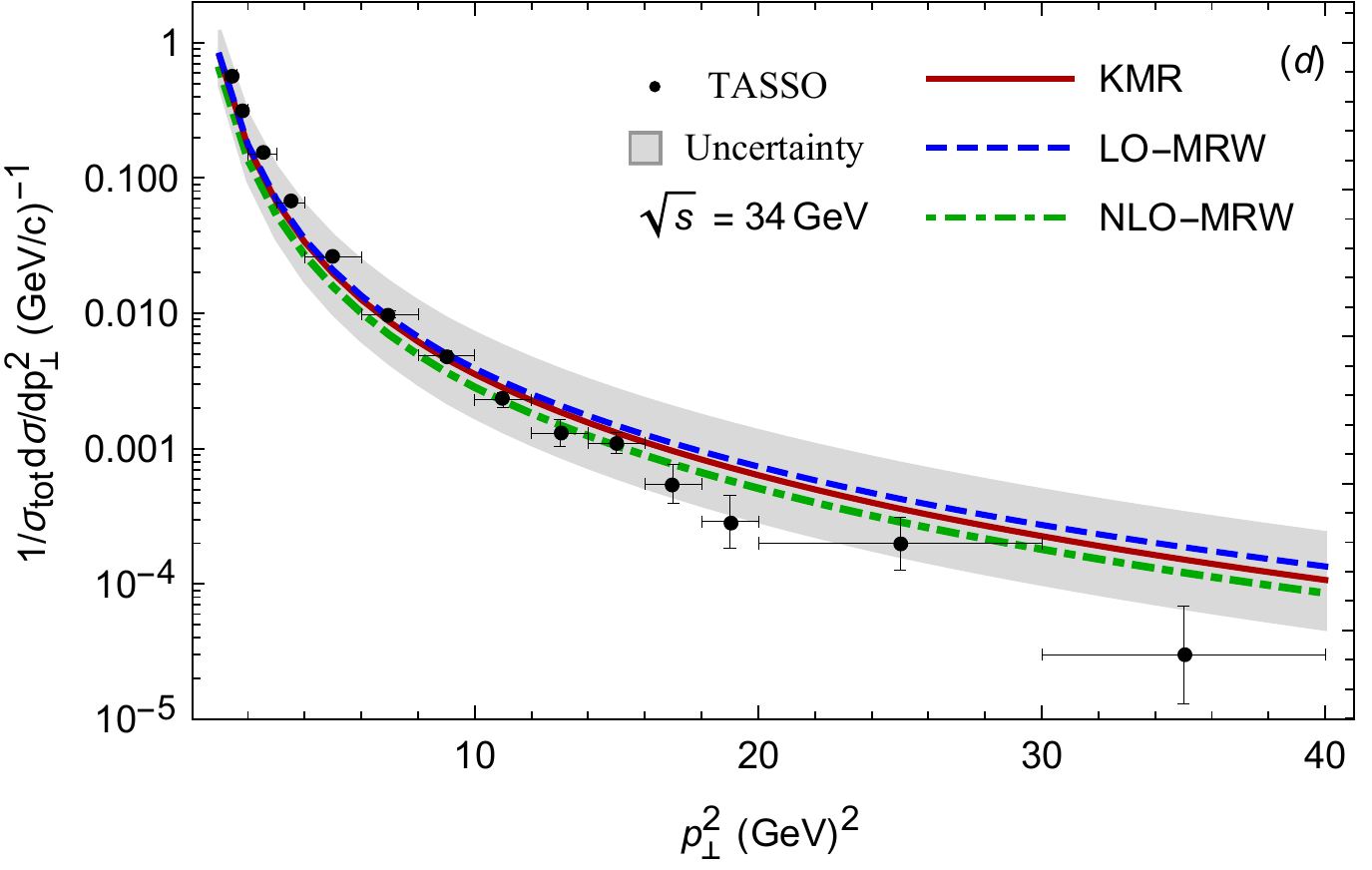}} \\
     \end{tabular}
\caption{The panel (a): the normalized distribution of the multiplicity with respect to   $p_\bot$ for charged particles is compared to the experimental data of CELLO \cite{CELLO}. The panels (b)-(d): the normalized differential cross sections with respect to   $p_\bot$ compared to the experimental data  of TASSO \cite{TASSO2} at the different CM energies. The shaded uncertainty grey bands are belong to the KMR prescription.\label{fig:2}}
      \end{center}
\end{figure*}

\begin{figure*}[htp]
  \begin{center}
    \begin{tabular}{cc}
      {\includegraphics[width=80mm,height=60mm]{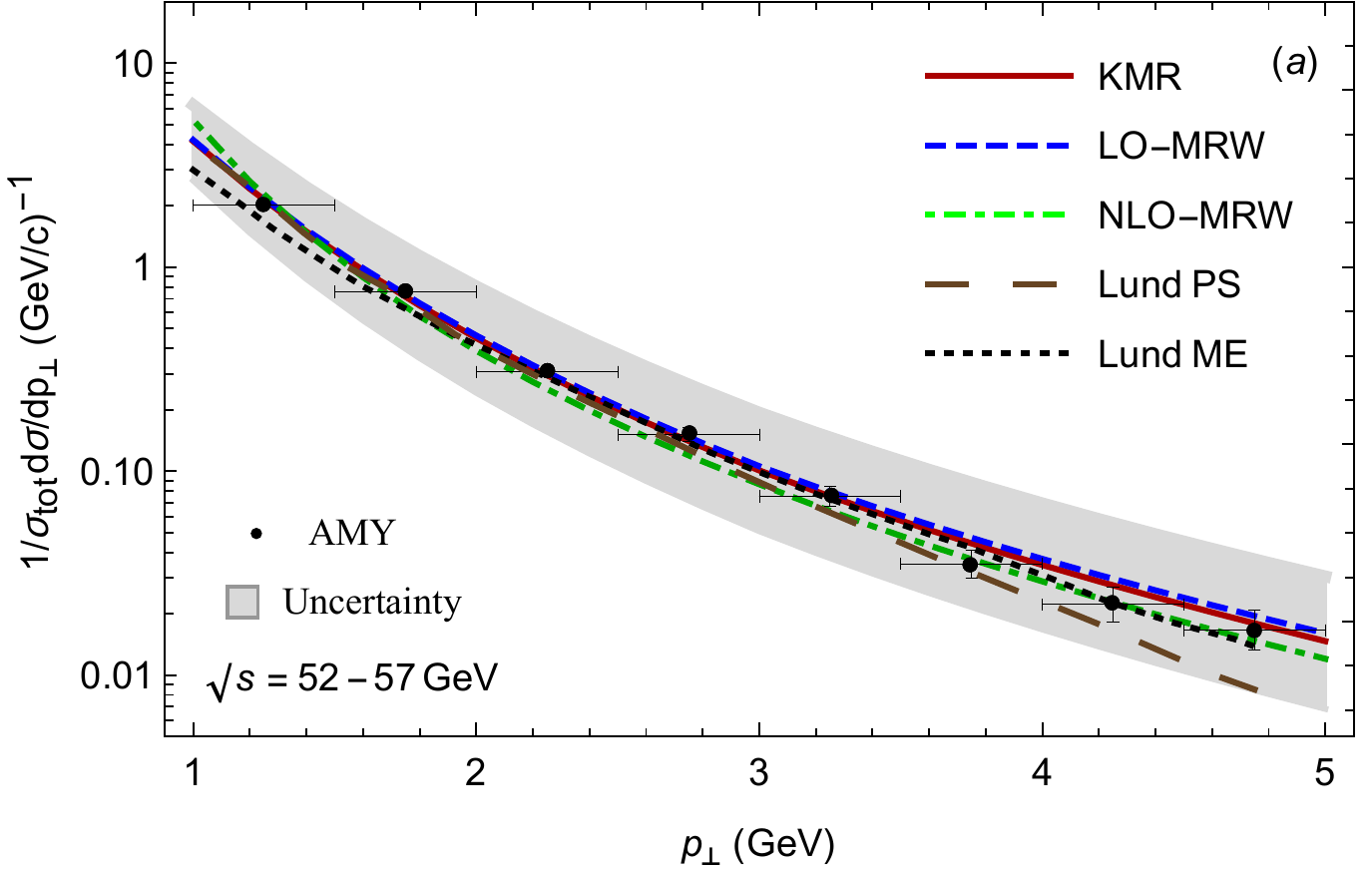}} &
       {\includegraphics[width=80mm,height=60mm]{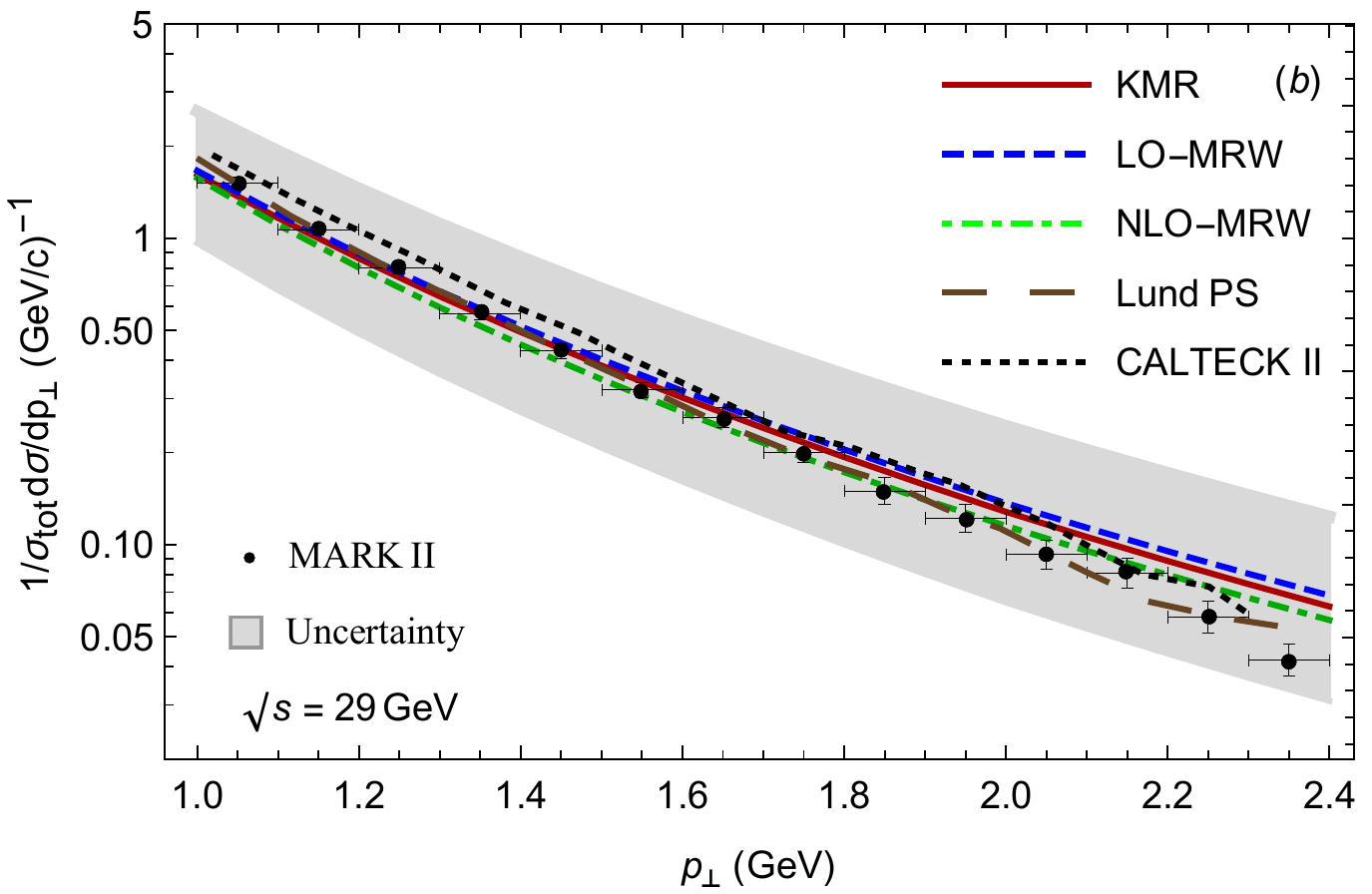}} \\
          \end{tabular}
\caption{The normalized differential cross sections with respect to   $p_\bot$ for charged particles is compared to the experimental data of AMY \cite{AMY} (the left panel),  of MARK II \cite{mark2} (the right panel), and   some "QCD+fragmentation" models predictions. The shaded uncertainty grey bands are belong to the KMR prescription.\label{fig:3}}
      \end{center}
\end{figure*}

\begin{figure*}[htp]
  \begin{center}
    \begin{tabular}{cc}
      {\includegraphics[width=80mm,height=60mm]{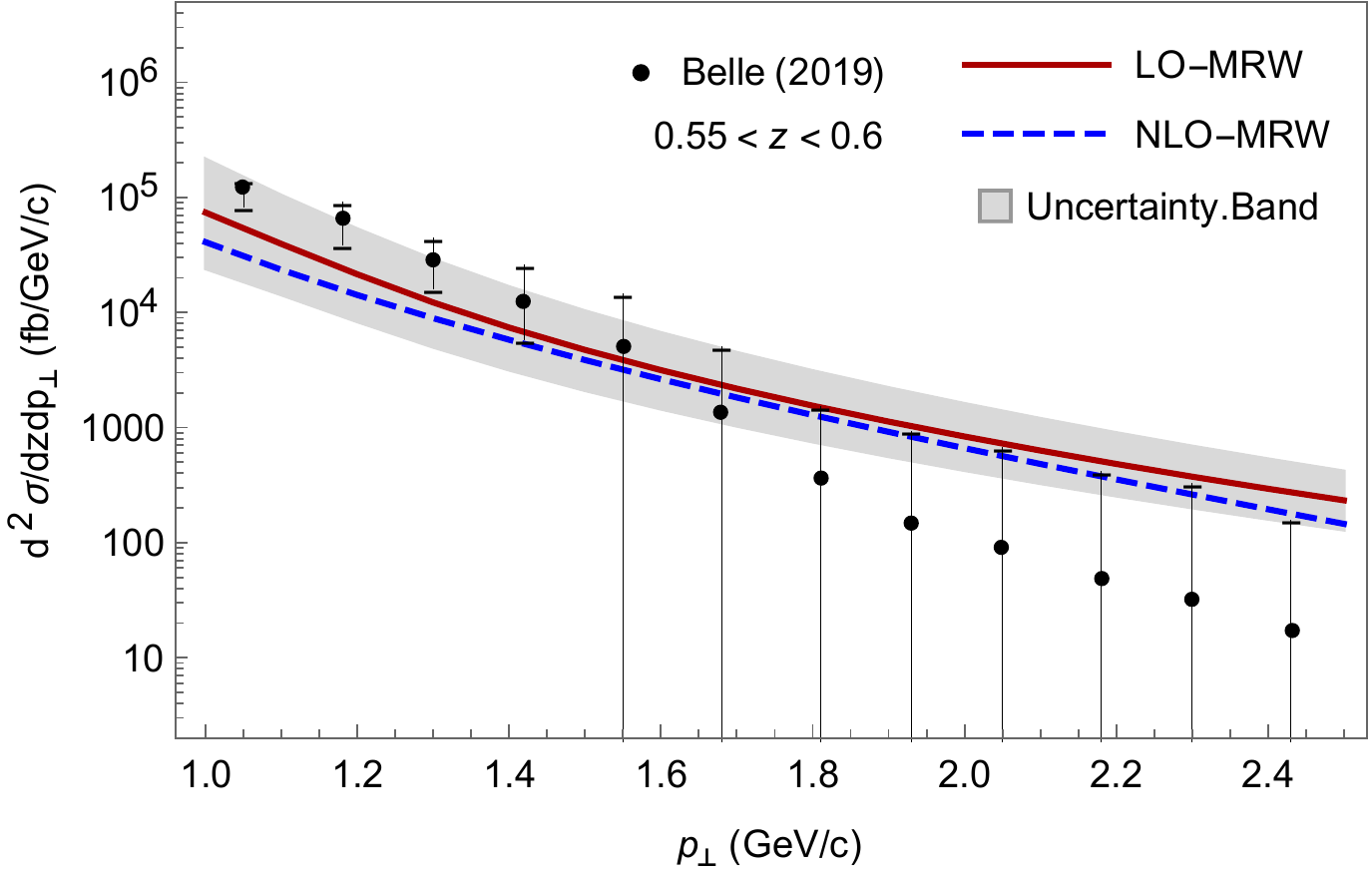}} &
       {\includegraphics[width=80mm,height=60mm]{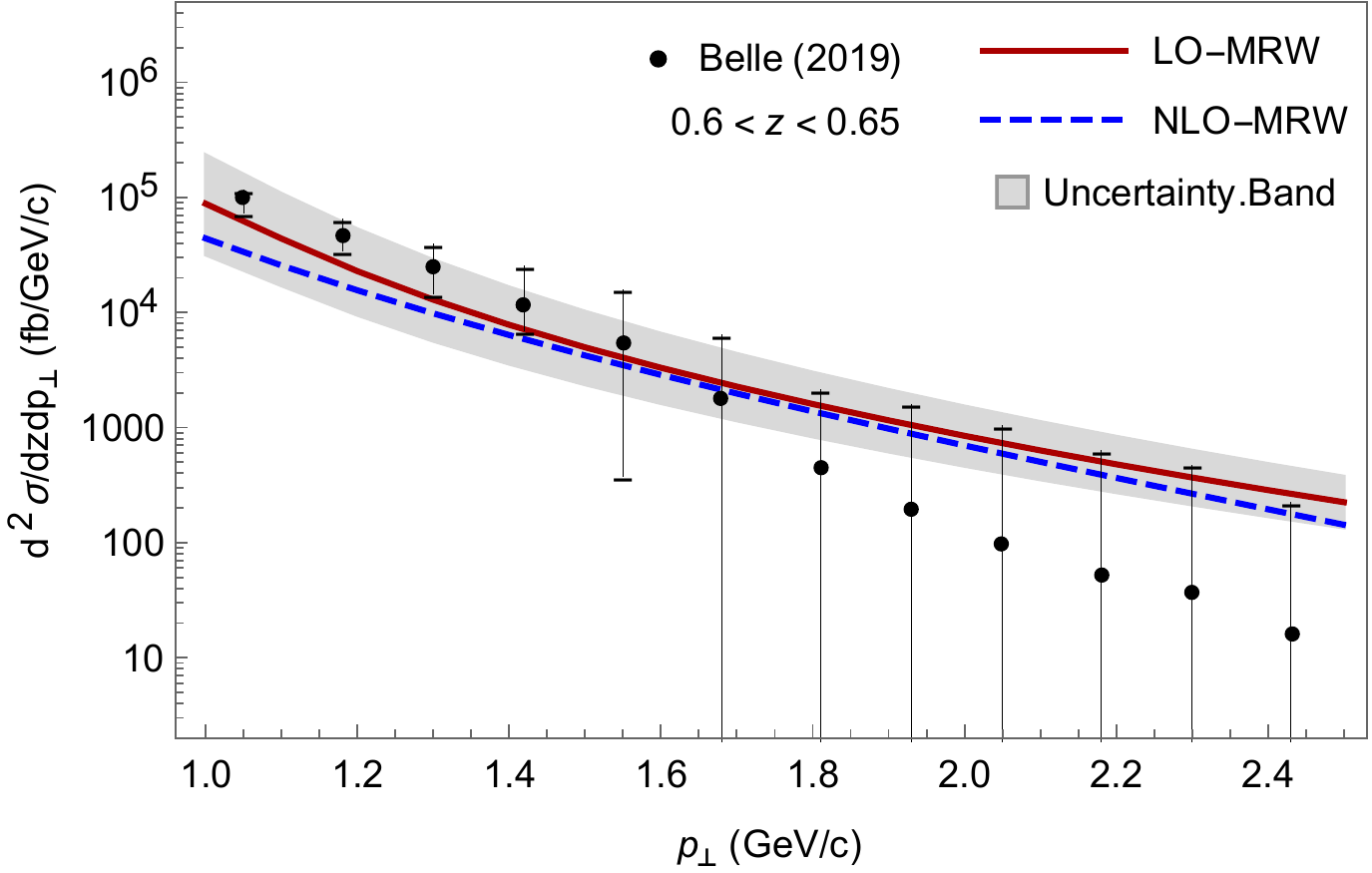}} \\
       {\includegraphics[width=80mm,height=60mm]{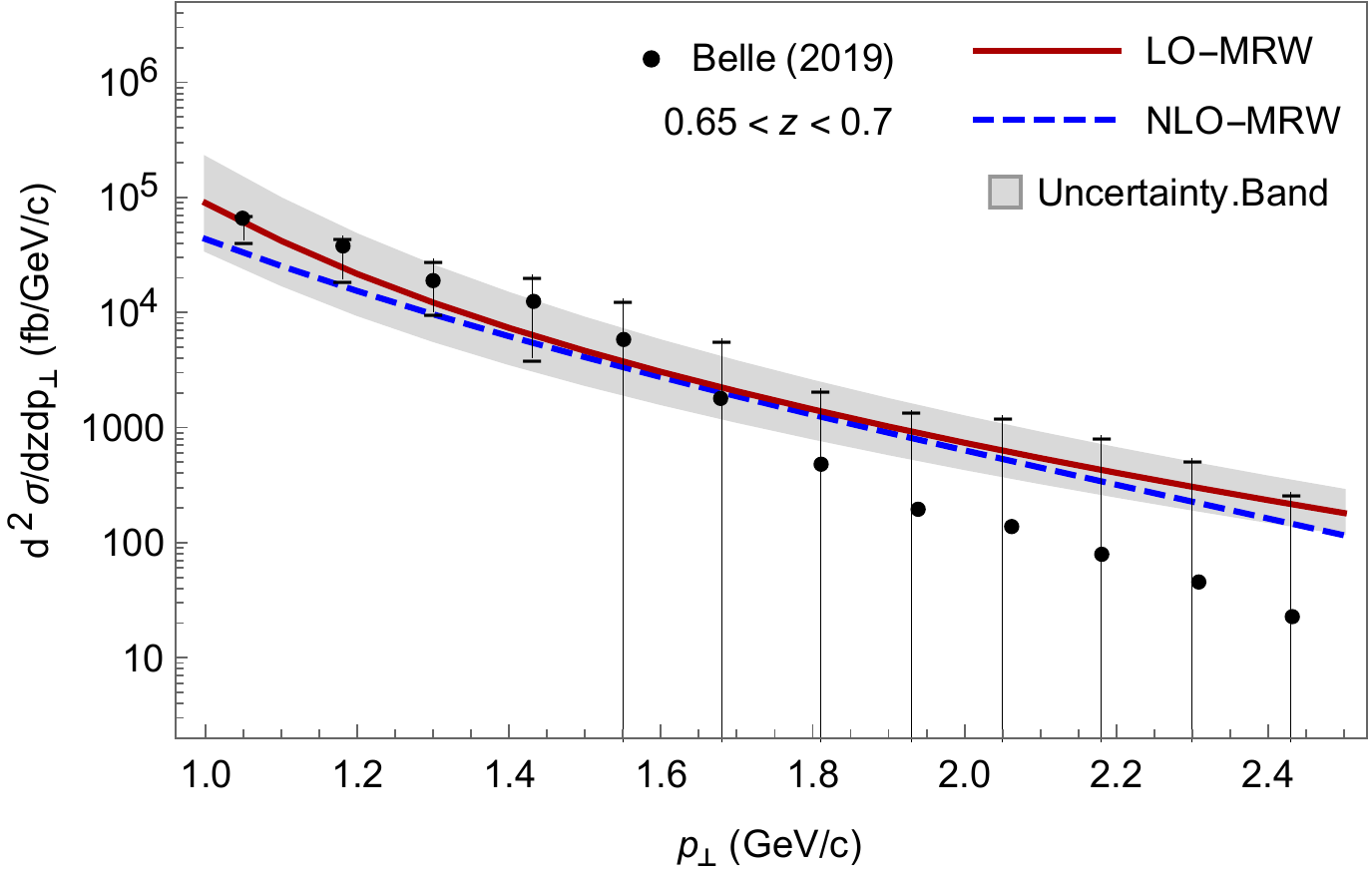}} &
       {\includegraphics[width=80mm,height=60mm]{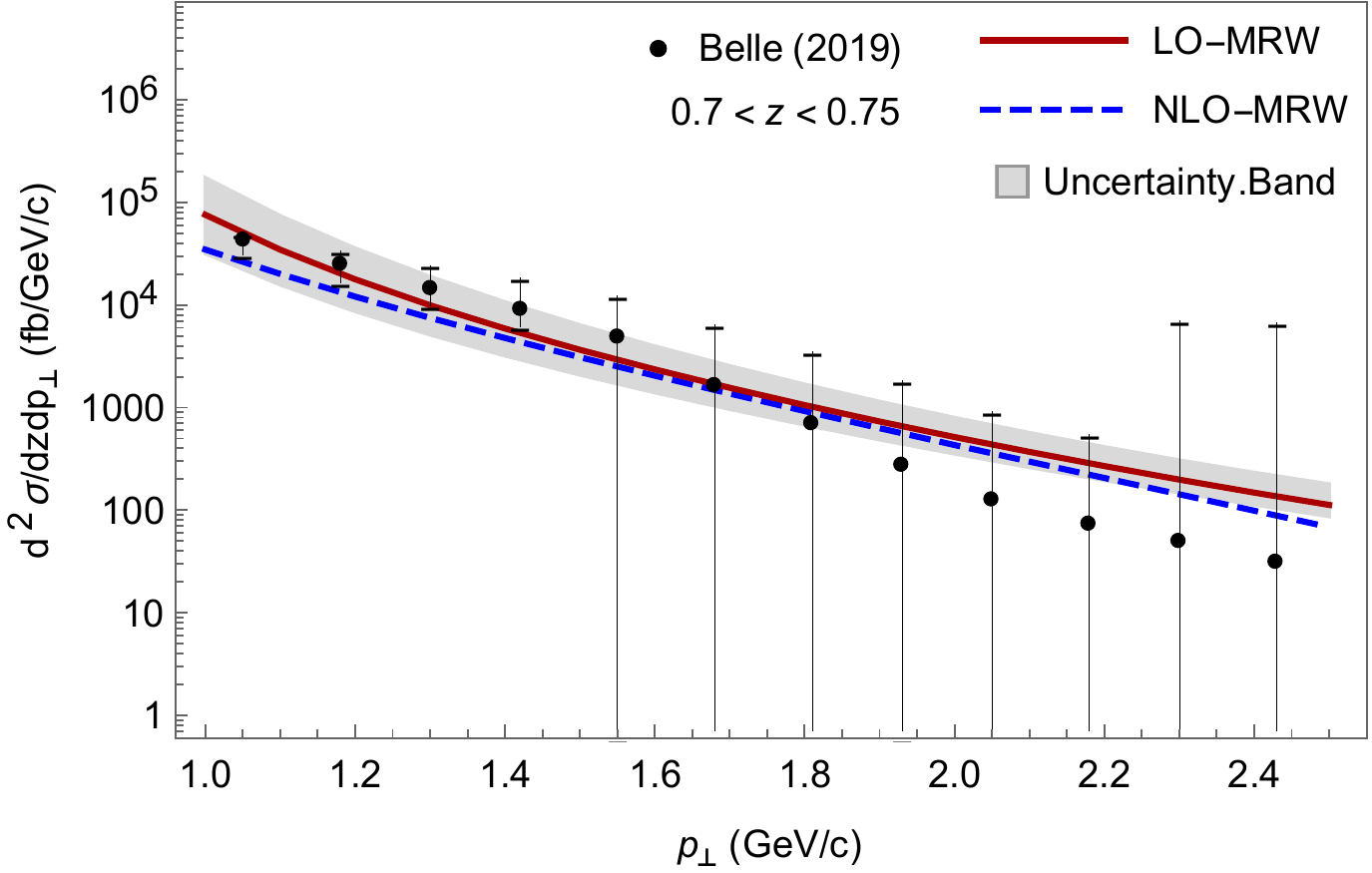}} \\
     \end{tabular}
\caption{The  differential cross sections for pions as a function of  $p_\bot$ for the indicated z bins and thrust $0.85 < T < 0.9$. The error  grey bands represent the uncertainties for LO-MRW formalism. The results are compared to the experimental data of Belle collaboration \cite{BR3} in the $\sqrt s=10.58$ GeV center of mass energy.  \label{fig:4}}
      \end{center}
\end{figure*}

\begin{figure*}[htp]
  \begin{center}
    \begin{tabular}{cc}
      {\includegraphics[width=80mm,height=60mm]{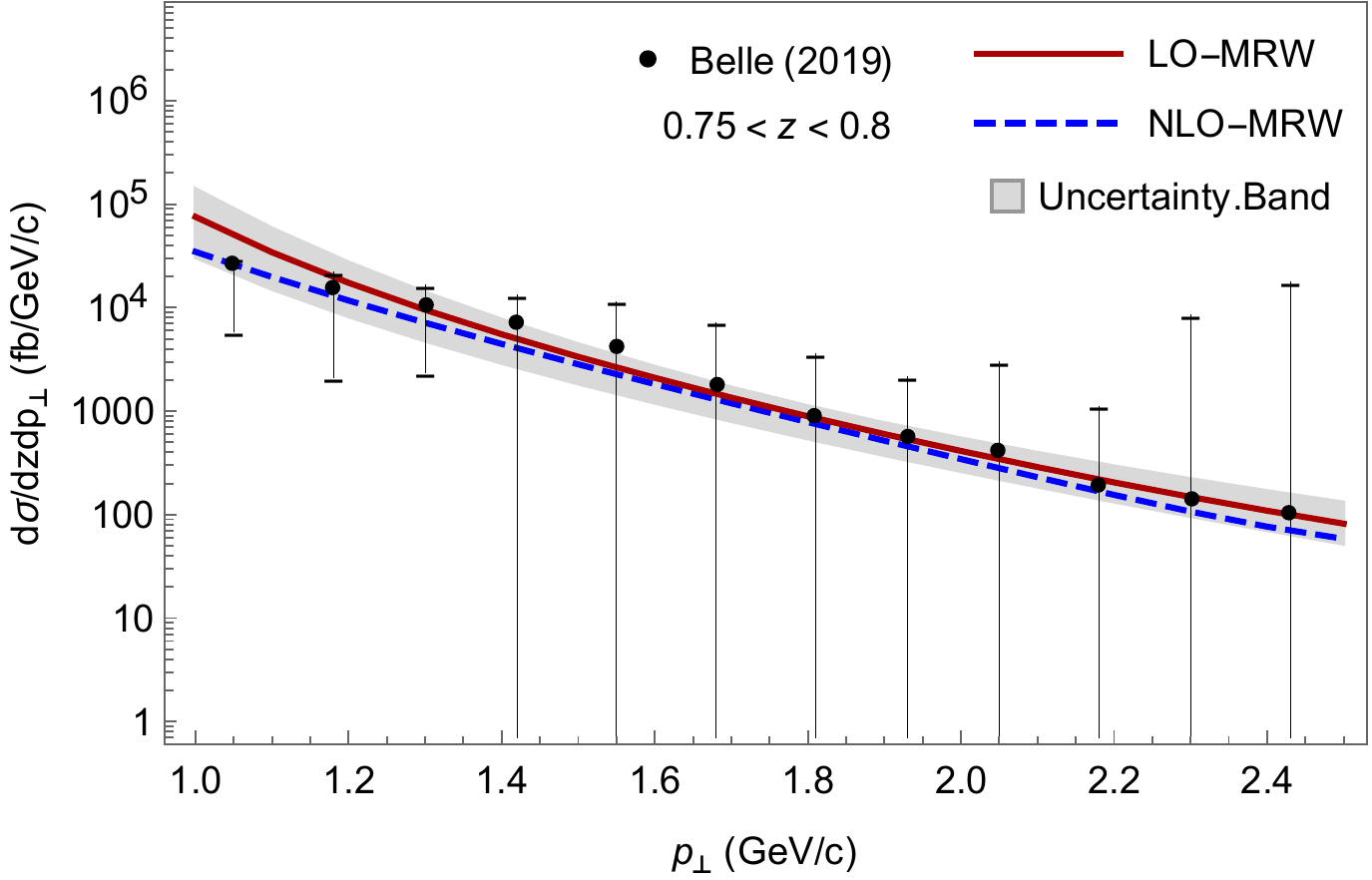}} &
       {\includegraphics[width=80mm,height=60mm]{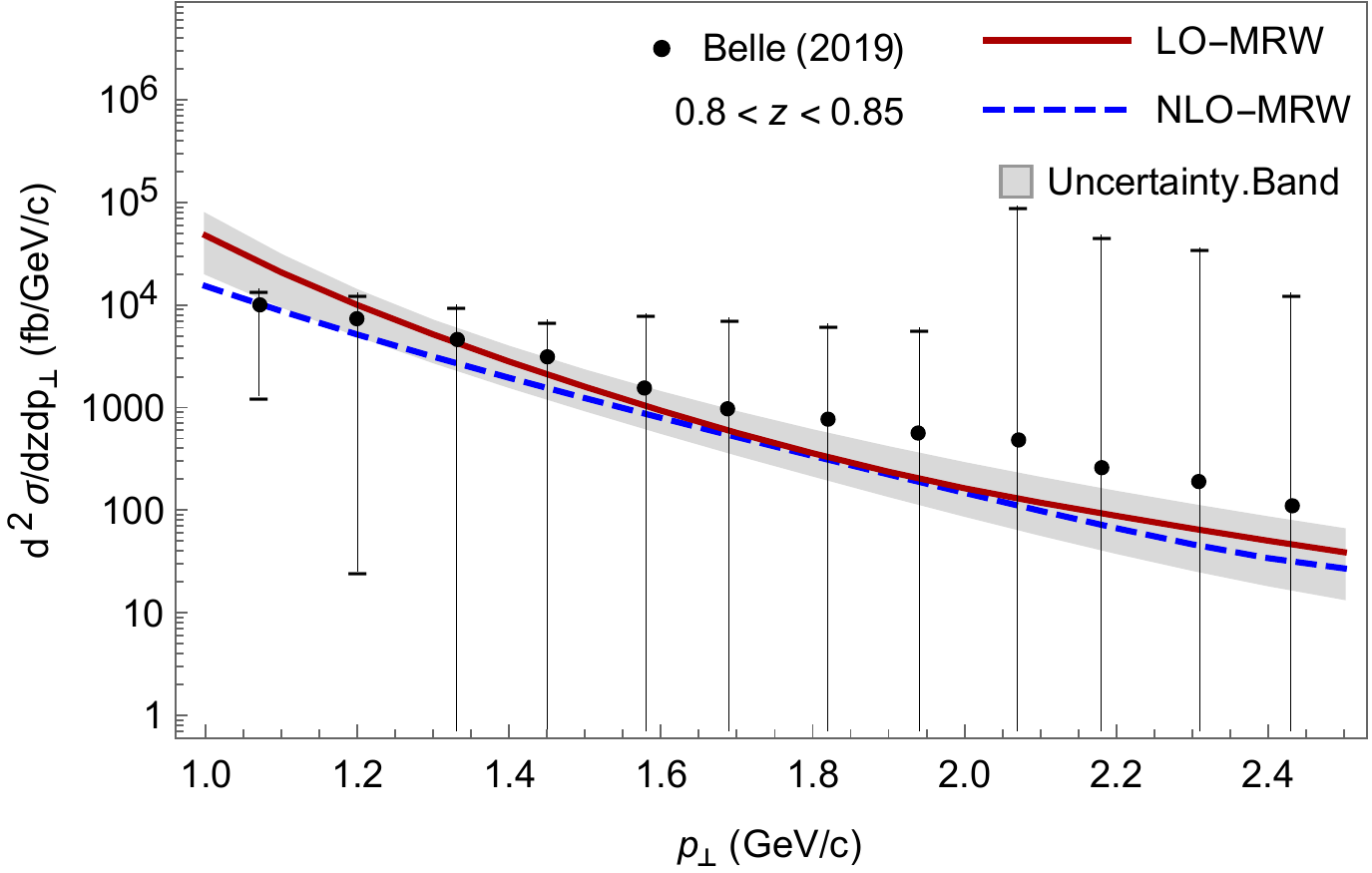}} \\
          \end{tabular}
\caption{ The same as the figure 5. \label{fig:5}}
      \end{center}
\end{figure*}

The results of the above numerical calculations are compared with the available
experimental data sets of the single inclusive hadron production in the ${e^ + }{e^ - }$ annihilation processes of TASSO detector at PETRA (DESY) \cite{TASSO1,TASSO2} laboratories and Belle detector at the KEKB \cite{BR3}. We use the data from different groups such as the AMY, MARK II, and CELLO collaborations. The results are
demonstrated in the figures \ref{fig:1}-\ref{fig:5} and compared with the different collaborations data at different CM energies.
In the figures figures \ref{fig:1}-\ref{fig:3}, the numerical results related to the UFF are shown by  the solid, dash, and dotted-dash
 curves correspond to the result of  different schemes namely the KMR, LO-MRW, and NLO-MRW, respectively.

We start by analyzing  the cross sections related
to the low CM energies for $\sqrt s  = 14, 22, 35$
and $44$ $GeV$. The results of the normalized differential cross sections
($1/\sigma_{tot}$)$d\sigma/dp_\bot$ with respect to   $p_\bot$ are compared to the experimental data of TASSO at the different CM energies which are shown in the figure \ref{fig:1}. It is demonstrated  that as the CM energy is increased the differential cross sections evaluated by using the three UFFs schemes,   become near to each other, especially in the case of  KMR and LO-MRW formalisms. On the other hand, our results become closer to the data for the lower values of $p_\bot$ and the higher values of CM energies.

In the   figure \ref{fig:2},  panel (a), the normalized distribution of the multiplicity with respect to   $p_\bot$ for charged particles:
\begin{equation}
{1 \over N}{dN \over dp_\bot}=2\pi {p_\bot}\frac{1}{{\sum\limits_q {{e_q}^2} }}\int {\sum\limits_q 
{{e_q}^2}{ [D_q^H(z,{p_\bot};{\mu^2}) + } D_{\bar q}^H(z,{p_\bot};{\mu^2})]dz}.
 \end{equation}
is compared to the experimental data of CELLO \cite{CELLO} while in the panels (b)-(d), the normalized differential cross sections
with respect to   $p_\bot^2$ are compared to the experimental data  of TASSO \cite{TASSO2} at the different CM energies.  The same conclusion can be made for this figure as the one we made for the figure \ref{fig:1}. It is also observed that the KMR and NLO-MRW are   closer to each other.

In the figure \ref{fig:3}, the normalized differential cross sections with respect to   $p_\bot$ for charged particles is compared to the experimental data of AMY \cite{AMY} (the panel (a)), and   MARK II \cite{mark2} (the panel (b)). A comparison between our results and some Monte Carlo techniques i.e., the Lund parton shower (Lund PS)(the dash lines in the both panels), the Lund matrix elements (Lund ME)(the dotted line in the  panel (a)) and the CALTECH II (the dotted line in the  panel (b)) models
 are presented. It is observed that
 our results are similar to those of  QCD+fragmentation function models. It seems that the Lund parton shower model provides a better description of data, since according to the reference \cite{AMY}, the total $\chi^2$ of the fits following this approach is the lowest among the three models. However, the advantage of the $k_t$-factorization methodology is that the calculation is completely perturbative and we do not use any fitting procedure to have a prediction of the data. 
 
In the figures 5 and 6, our results of LO- and NLO-MRW are compared with the differential cross section data sets of pions from the Belle collaboration \cite{BR3} as a function of transverse momentum $p_\bot$ for the indicated $z$ bins and thrust value $0.85 < T < 0.9$. These figures show as the amount of $z$ bin increases, the result of NLO-MRW scheme becomes closer to the data. 

As we pointed out before, it is obvious that all of the three approaches have similar  behavior. Although  near   $p_\bot\sim1$, there is not any significant preference between the results of three schemes, but by increasing $p_\bot$  they start to separate from  each other. According to these panels, our results show a bit underestimate and overestimate in the low and high $p_\bot$ region. However, the uncertainty bands of our calculations cover the experimental data. Thus, one can conclude that our perturbative and straightforward calculations (using the KMR and MRW methods) give a good description of the data. 

In different panels of the figure 7, we compare our differential cross section defined in the equation (7), using the LO-MRW formalism,  with those of new Pythia 6.4 and Pythia 8.2 parton showers \cite{R2 new 2} as well as  the DELPHI \cite{R2 new 3}, SLD
\cite{R2 new 4} and ALEPH \cite{R2 new 5} collaboration data, at CM energy 91 GeV. There is good agreement between our prediction and the mentioned experimental data as well as new parton showers, specialty for the  Pythia 8.2 parton shower \cite{R2 new 2}.

 \begin{figure*}[htp]
  \begin{center}
    \begin{tabular}{cc}
      {\includegraphics[width=80mm,height=60mm]{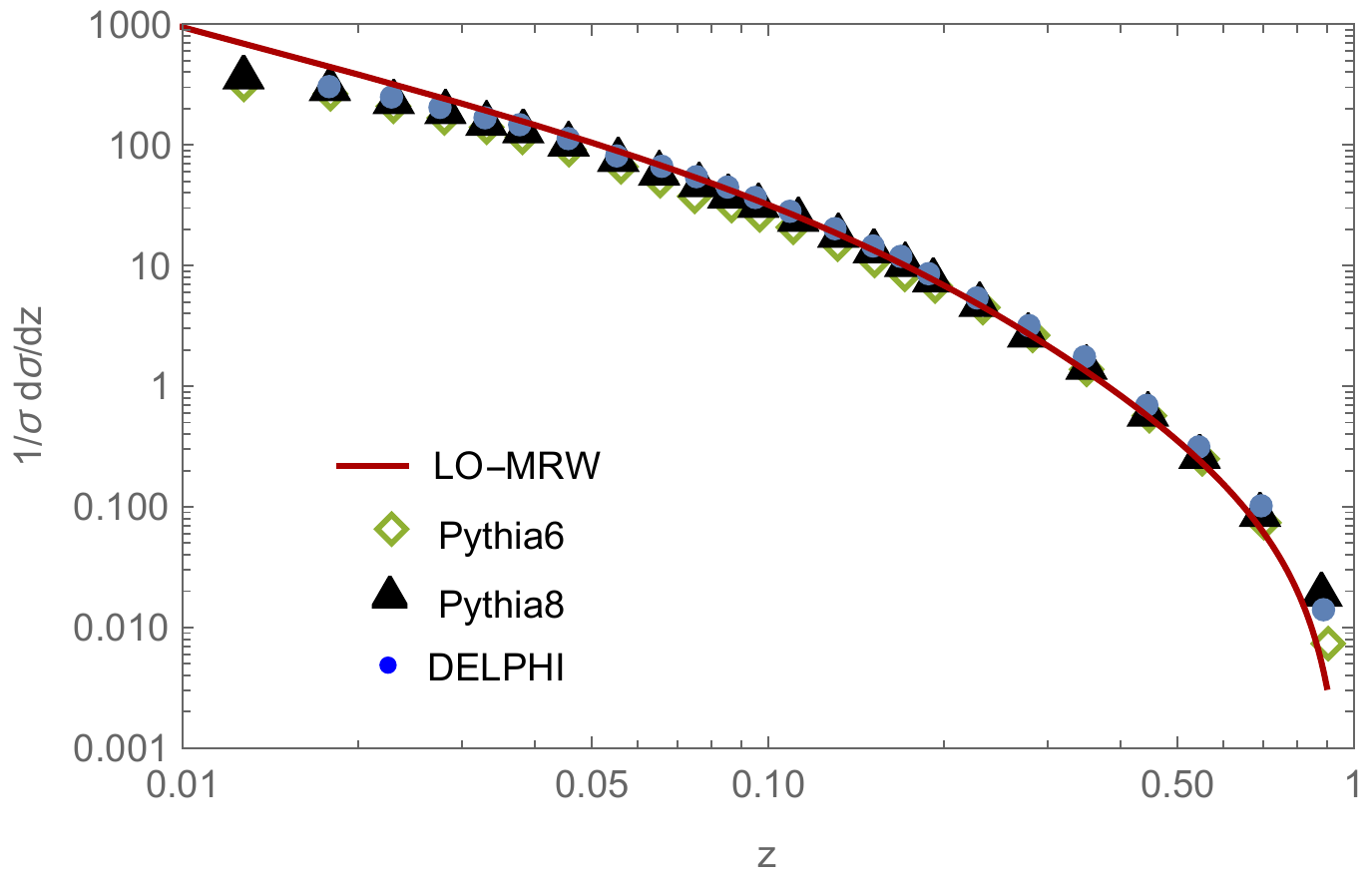}} &
       {\includegraphics[width=80mm,height=60mm]{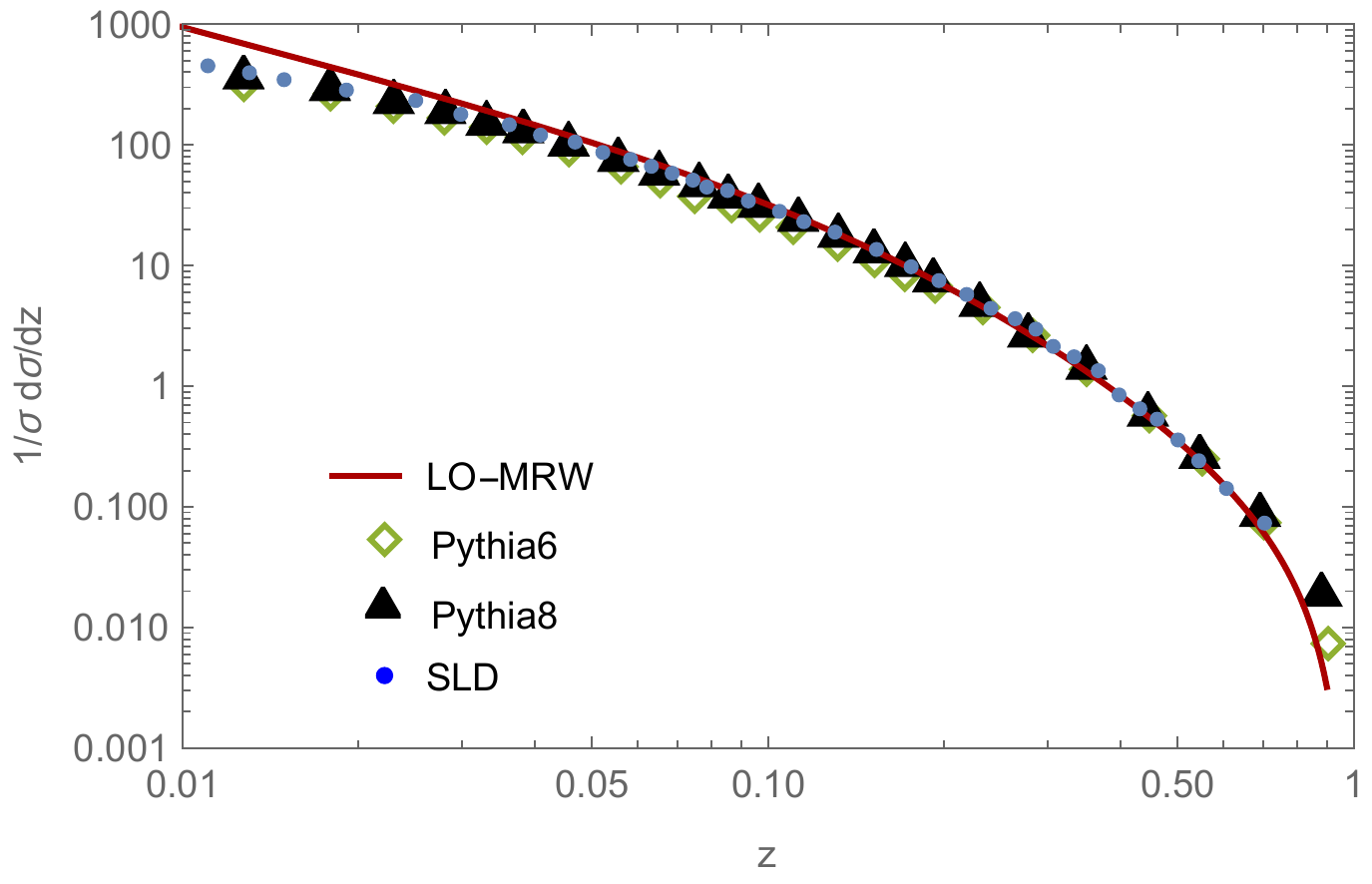}} \\
       {\includegraphics[width=80mm,height=60mm]{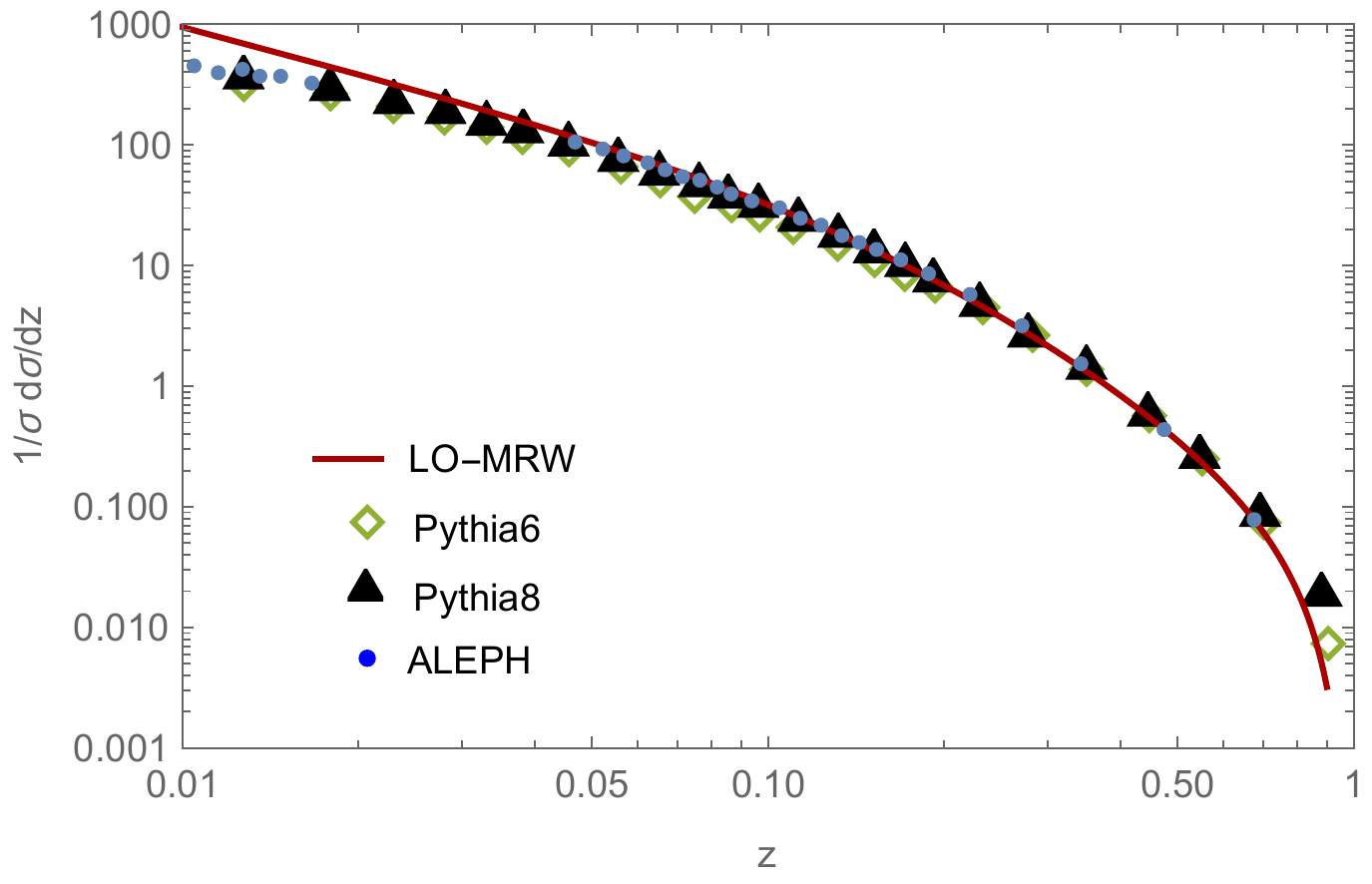}} &
          \end{tabular}
\caption{ The comparison  of our differential cross section defined in the equation (7), using the LO-MRW formalism,  with those of Pythia 6.4 and Pythia 8.2 parton showers \cite{R2 new 2} as well as the DELPHI \cite{R2 new 3}, SLD
\cite{R2 new 4} and ALEPH \cite{R2 new 5}  collaborations data, at CM energy 91 GeV. 
\label{fig:6}}
\end{center}
\end{figure*}
\section{Conclusions}
We presented the first analysis of the applicability of the $k_t$-factorization approach in the single inclusive hadron production in the ${e^ + }{e^ - }$ annihilation processes. We used the transverse momentum dependent fragmentation functions of three different prescriptions, i.e., KMR,
MRW, and NLO-MRW.  We
calculated several distributions of normalized transverse momentum and multiplicity of the charged fragmented hadrons in the leading order. In addition, we
obtained the uncertainty band for the cross section distribution in
the case of KMR by changing the scale factor as illustrated in the section
III. We found with a good approximation, all three schemes give a similar and acceptable description of
data presented in this report.

In the reference \cite{R2 new 1} the transverse-momentum-dependent FFs (or UFFs) up to N3LO QCD were calculated. However,  it is the first time that present formalism is applied to calculate the unintegrated fragmentation functions up to the NLO level and should be considers as the first step for application of KMR and MRW formalisms. On the other hand  we only  considered the lowest order structure functions for the evaluation of differential cross sections. So, we hope by extending our formalism to the next-leading order in the structure functions and improving of the UFFs, we could get better accuracy, in the future works.
\begin{acknowledgements}
M. Modarres and R. Taghavi would like to acknowledge the research
support of the University of Tehran and   the Iran National Science
Foundation (INSF) for their  grants.
\end{acknowledgements}

\appendix
\section{  }
\label{b} The NLO splitting functions are defined as \cite{pab}:

    \begin{equation}
\tilde{P}_{ab}^{(0+1)}(z) = \tilde{P}_{ab}^{(0)}(z) + {\alpha_S
\over 2\pi}
    \tilde{P}_{ab}^{(1)}(z),
    \label{eq12p}
    \end{equation}
with
    \begin{equation}
\tilde{P}_{ab}^{(i)}(z) = P_{ab}^{(i)}(z) - \Theta (z-(1-\Delta))
\delta_{ab} F^{(i)}_{ab} P_{ab}(z),
    \label{eq13}
    \end{equation}
where $i= 0$ and $1$ stand for the $LO$ and the $NLO$, respectively.
$\Delta$ can be defined as \cite{MRW}:
    $$ \Delta = {\kappa_t\sqrt{1-z} \over \kappa_t\sqrt{1-z} + \mu}.$$ and we have:
\begin{equation}
F_{qq}^{(0)} = {C_F},
 \label{eq12pp}
    \end{equation}
\begin{equation}
F_{qq}^{(1)} =  - {C_F}({T_R}{N_F}\frac{{10}}{9} + {C_A}(\frac{{{\pi
^2}}}{6} - \frac{{67}}{{18}})),
 \label{eq12ppp}
    \end{equation}

\begin{equation}
F_{gg}^{(0)} = 2{C_A},
 \label{eq12pppp}
    \end{equation}

\begin{equation}
F_{gg}^{(1)} =  - 2{C_F}({T_R}{N_F}\frac{{10}}{9} +
{C_A}(\frac{{{\pi ^2}}}{6} - \frac{{67}}{{18}})),
 \label{eq12ppppp}
    \end{equation}
\begin{equation}
{P_{qq}}(z) = \frac{{(1 - {z^2})}}{{1 - z}},
 \label{eq12pppppppp}
    \end{equation}
\begin{equation}
{P_{gg}}(z) = \frac{z}{{(1 - z)}} + \frac{{(1 - z)}}{z} + z(1 - z),
 \label{eq127}
    \end{equation}

\newpage


\begin{thebibliography}{a}
\addcontentsline{toc}{chapter}{Bibliographie}
\bibitem{DGLAP1} V.N. Gribov and L.N. Lipatov, Yad.Fiz., 15 (1972) 781.                                                                                   
\bibitem{DGLAP2} L.N. Lipatov, Sov.J.Nucl.Phys., 20 (1975) 94.                                                                                              
\bibitem{DGLAP3} G. Altarelli and G. Parisi, Nucl.Phys.B, 126 (1977) 298.                                                                                
\bibitem{DGLAP4} Y.L. Dokshitzer, Sov.Phys.JETP, 46 (1977) 641.  
\bibitem{aybat} Aybat S.M. and Rogers T.C., Phys. Rev., D, 83 (2011) 114042.                                                                                        
\bibitem{lie}M. Lie, B. M,  Phys. Rev. 98  (2018) 036024.
\bibitem{mypaper2}R. Taghavi, M. Mirjalili, Modern Physics Letters A, 32, No. 10 (2017) 1750040.
\bibitem{KMR} M.A. Kimber, A.D. Martin, M.G. Ryskin, Phys.Rev.D, 63 (2001) 114027.  
\bibitem{MRW}A.D. Martin, M.G. Ryskin, G. Watt, Eur.Phys.J.C, 66 (2010) 163.     
\bibitem{10}M. Modarres, H. Hosseinkhani, N. Olanj, Nucl.Phys.A, 902 (2013) 21.                                                             
\bibitem{11} M. Modarres, H. Hosseinkhani, Few-Body Syst., 47 (2010) 237.                                                                        
\bibitem{12} M. Modarres, H. Hosseinkhani, Nucl.Phys.A, 815 (2009) 40.                                                               
\bibitem{13} H. Hosseinkhani, M. Modarres, Phys.Lett.B, 694 (2011) 355.                                                                             
\bibitem{14} H. Hosseinkhani, M. Modarres, Phys.Lett.B, 708 (2012) 75. 
\bibitem{OWENS1} J. F. Qwens, Phys. Lett. B, 76 (1978) 1.
\bibitem{Owens1}J.F. Owens, Phys.Rev.D,  20 (1979) 221. 
\bibitem{Owens2}J.F. Owens, Rev.Mod.Phys., 59 (1987) 465.
\bibitem{FF1}S. Albino, Rev. Mod. Phys. 82, 2489 (2010).
\bibitem{Metz} A. Metz, A. Vossen,  Prog. Part. Nucl. Phys. 91 (2016) 136.
\bibitem{clean1} J. C. Collins and D. E. Soper, Nucl. Phys. B, 194 (1982) 445.
\bibitem{BABAR1}J. P. Lees et al. (BABAR Collaboration), Phys. Rev. D, 90 (2014) 052003.
\bibitem{BABAR2} J. P. Lees et al.(BABAR Collaboration), Phys. Rev. D, 92 (2015) 111101.
\bibitem{collins1}Z.-B. Kang, A. Prokudin, P. Sun, F. Yuan,Phys. Rev. D, 91 (2015) 071501.
\bibitem{collins2}A. Bacchetta, M.G. Echevarria, P.J.G. Mulders, M. Radici, A. Signori,  J. High Energy Phys. 11 (2015) 076.
\bibitem{collins4}M. Anselmino, M. Boglione, U. D’Alesio, J.O. Gonzalez Hernandez, S. Melis, F. Murgia, A. Prokudin, Phys. Rev. D, 93(3) (2016) 034025.
\bibitem{f1}A. Signori, A. Bacchetta, M. Radici, G. Schnell, J. High Energy Phys. 1311 (2013) 194.
\bibitem{f2}M. Anselmino, M. Boglione, J. Gonzalez Hernandez, S. Melis, A. Prokudin,  J. High Energy Phys. 1404 (2014) 005.
\bibitem{f3}C.A. Aidala, B. Field, L.P. Gamberg, T.C. Rogers, Phys. Rev. D, 89(9) (2014) 094002.
\bibitem{f4}J. Collins, L. Gamberg, A. Prokudin, T.C. Rogers, N. Sato, B. Wang,  Phys. Rev. D, 94(3) (2016) 034014.
\bibitem{f5}M. Boglione, J. Collins, L. Gamberg, J.O. Gonzalez-Hernandez, T.C. Rogers, N. Sato, Phys. Lett. B, 766 (2017) 245.
\bibitem{mypaper1}M. Boglione, J.O. Gonzalez-Hernandez, R. Taghavi, Phy.Lett.B, 772 (2017) 78.  
 \bibitem{BR1} Alessandro Bacchetta, Filippo Delcarro, Cristian Pisano, Marco Radicib
and Andrea Signoric, JHEP, 06 (2017) 081.
\bibitem{BR2} Zhong-Bo Kang, Ding Yu Shaoa and Fanyi Zhaoa, JHEP, 12 (2020) 127.
\bibitem{f7}Belle Collaboration, Phys. Rev. D, 99 (2019) 112006.              
\bibitem{TASSO1}M. Althoff, et al., Z. Phys. C, 22 (1984) 307 .
\bibitem{TASSO2}W. Braunschweig, et al., Z. Phys. C, 47 (1990) 187.
\bibitem{mark2}A. Petersen, et al., Phys. Rev. D, 37 (1988) 1.
\bibitem{AMY} AMY collaboration, Phys. Rev. D, 41 (1990) 9.
\bibitem{CELLO} CELLO collaboration, Z.Phys.C, 14 (1982) 189 .
\bibitem{R2 new 2}H. Bello Martínez, R. J. Hernández-Pinto and I. León Monzón, 7th Annual Conference on Large Hadron Collider Physics - LHCP2019
20-25 May, (2019), arXiv:1910.03035v1 [hep-ph].
\bibitem{R2 new 3}  DELPHI Collaboration, Eur.Phys.J.C,   05   (1998) 585.
\bibitem{R2 new 4} SLD Collaboration, Phys.Rev.D, 59 (1999)  052001.
 \bibitem{R2 new 5} ALEPH Collaboration, Z.Phys.C 66 (1995) 355.
\bibitem{BR3} Belle Collaboration, R. Seidl, et al, Phys.Rev.D, 99 (2019)  112006.
\bibitem{LundPS} T. Sjostrand and M. Bengtsson, Comput. Phys. Commun. 43 (1987) 367.    
\bibitem{LundME}F. Gutbord, G. Kramer, and G. Schierholz, Z. Phys. C, 21 (1984) 235.   
\bibitem{caltech2}T. D. Gottschalk and D. Morris, Nucl. Phys. B, 288 (1987) 729.                                              
\bibitem{Altarelli} G. Altarelli, Phys.Rep., 81 (1982) 1. 
\bibitem{f1fh}D. Florian, M. Stratmann, and W. Vogelsang, Phys. Rev. D, 57 (1998) 5811.
\bibitem{CCFM1} M. Ciafaloni, Nucl.Phys.B, 296 (1988) 49.                                                                                                      
\bibitem{CCFM2} S. Catani, F. Fiorani, and G. Marchesini, Phys.Lett.B, 234 (1990) 339.                                                        
\bibitem{CCFM3} S. Catani, F. Fiorani, and G. Marchesini, Nucl.Phys.B, 336 (1990) 18.                                                           
\bibitem{CCFM4} M. G. Marchesini, Proceedings of the Workshop QCD at 200 TeV Erice,
 Italy, edited by L. Cifarelli and Yu.L. Dokshitzer, Plenum, New York (1992) 183.                                                                      
\bibitem{CCFM5} G. Marchesini, Nucl.Phys.B, 445 (1995) 49.              
\bibitem{44}G. Marchesini, B.R. Webber, Nucl.Phys.B, 310 (1988) 461.                                                                                 
\bibitem{45}Y.L. Dokshitzer, V.A.Khoze, S.I. Troyan, A.H. Mueller, Rev.Mod.Phys., 60 (1988) 373.
\bibitem{DSS}D. de Florian, R. Sassot, M. Stratmann, Phys. Rev. D, 76 (2007) 074033.      
\bibitem{pab}W. Furmanski and R. Petronzio, Phys. Lett. B, 97 (1980) 437. 
\bibitem{R2 new 1}Ming-xing Luo, Tong-Zhi Yang, Hua Xing Zhua and Yu Jiao Zhua, JHEP 06, 115 (2021).
                   \end{thebibliography}
\end{document}